\documentclass[english,preprint,showpacs,pra,superscriptaddress]{revtex4-1}
\usepackage[T1]{fontenc}
\usepackage[latin9]{inputenc}
\usepackage{color}
\usepackage{amsmath}
\usepackage{graphicx}

\makeatletter

\providecommand{\tabularnewline}{\\}

\usepackage{babel}

\makeatother

\usepackage{babel}
\begin{document}

\title{Memory-assisted quantum key distribution resilient against multiple-excitation
effects}

\author{Nicol$\grave{\rm o}$ Lo Piparo}

\affiliation{School of Electronic and Electrical Engineering, University of Leeds,
Leeds, UK}

\author{Neil Sinclair}

\affiliation{Institute for Quantum Science and Technology, and Department of Physics
\& Astronomy, University of Calgary, Calgary, Alberta T2N 1N4, Canada}

\author{Mohsen Razavi}

\affiliation{School of Electronic and Electrical Engineering, University of Leeds,
Leeds, UK}
\begin{abstract}
Memory-assisted quantum key distribution (MA-QKD) has recently been
proposed as a technique to improve the rate-versus-distance behavior
of QKD systems by using existing, or nearly-achievable, quantum technologies.
The promise is that MA-QKD would require less demanding quantum memories
than the ones needed for probabilistic quantum repeaters. Nevertheless,
early investigations suggest that, in order to beat the conventional
no-memory QKD schemes, the quantum memories used in the MA-QKD protocols
must have high bandwidth-storage products and short interaction times.
Among different types of quantum memories, ensemble-based memories
offer some of the required specifications, but they typically suffer
from multiple excitation effects. To avoid the latter issue, in this
paper, we propose two new variants of MA-QKD both relying on single-photon
sources (SPSs) for entangling purposes. One is based on known techniques
for entanglement distribution in quantum repeaters. This scheme turns
out to offer no advantage even if one uses ideal SPSs. By finding the
root cause of the problem, we then propose another setup, which can
outperform single no-QM setups even if we allow for some imperfections
in our SPSs. For such a scheme, we compare the key rate for different
types of ensemble-based memories and show that certain classes of
atomic ensembles can improve the rate-versus-distance behavior. 
\end{abstract}
\maketitle

\section{Introduction}

Providing secure key exchange at long distances is a yet-to-be
achieved objective for quantum key distribution (QKD) systems. While
some recent demonstrations have managed to exchange secret keys at
307~km \cite{307km} and 404~km \cite{MDI400km}, the key rate achieved
at such distances is extremely low. The limitation in going to further
distances is dictated by the exponentially-growing loss factor in
optical fibers \cite{SP}. Probabilistic quantum repeaters offer a
solution to extend the communication distance to over thousands of
kilometers \cite{DLCZ,IEEE2,Rate-loss_repeater,repeaters_review_2011}. However, such quantum
repeaters rely on quantum memory (QM) modules \cite{Khabat_review} with characteristics
that are hard to achieve with the current technology. This does not
necessarily mean that the existing QMs cannot offer any advantages.
In fact, it has been shown that by using imperfect memories in measurement-device-independent QKD (MDI-QKD) systems, one may beat the no-QM QKD systems in rate and range to enable inter-city
QKD operation \cite{Brus:MDIQKD-QM_2013,panayi}. Although, unlike
quantum repeaters, they are not scalable. Nonetheless, such memory-assisted (MA)
MDI-QKD setups can relax some of the demanding constraints on QMs,
leading to more feasible implementations. Early investigations
suggest that QMs with large storage-bandwidth products as well as
short access and entangling times are necessary for MA-MDI-QKD \cite{panayi}. These requirements may be achieved by QMs based on atomic ensembles \cite{Khabat_review}, with the added benefit of strong light-matter coupling, of which offers the possibility for efficient implementations. Ensemble-based QMs may, however, allow for storage of multiple excitations \cite{squ_state},
which have been shown to be deleterious to their performance \cite{IEEE1}. Here, we propose two MA-MDI-QKD schemes, both
relying on single-photon sources (SPSs), in an attempt to rectify
the multiple-excitation problem. We begin with what seems to be the
more obvious choice for our setup, but that turns out to not fully
solve the problem, even if one uses ideal SPSs. We then fix the problem
in our second setup, and show that it can outperform no-QM systems
even if the employed SPSs are not ideal.

\begin{figure}
\begin{centering}
\includegraphics[scale=0.6]{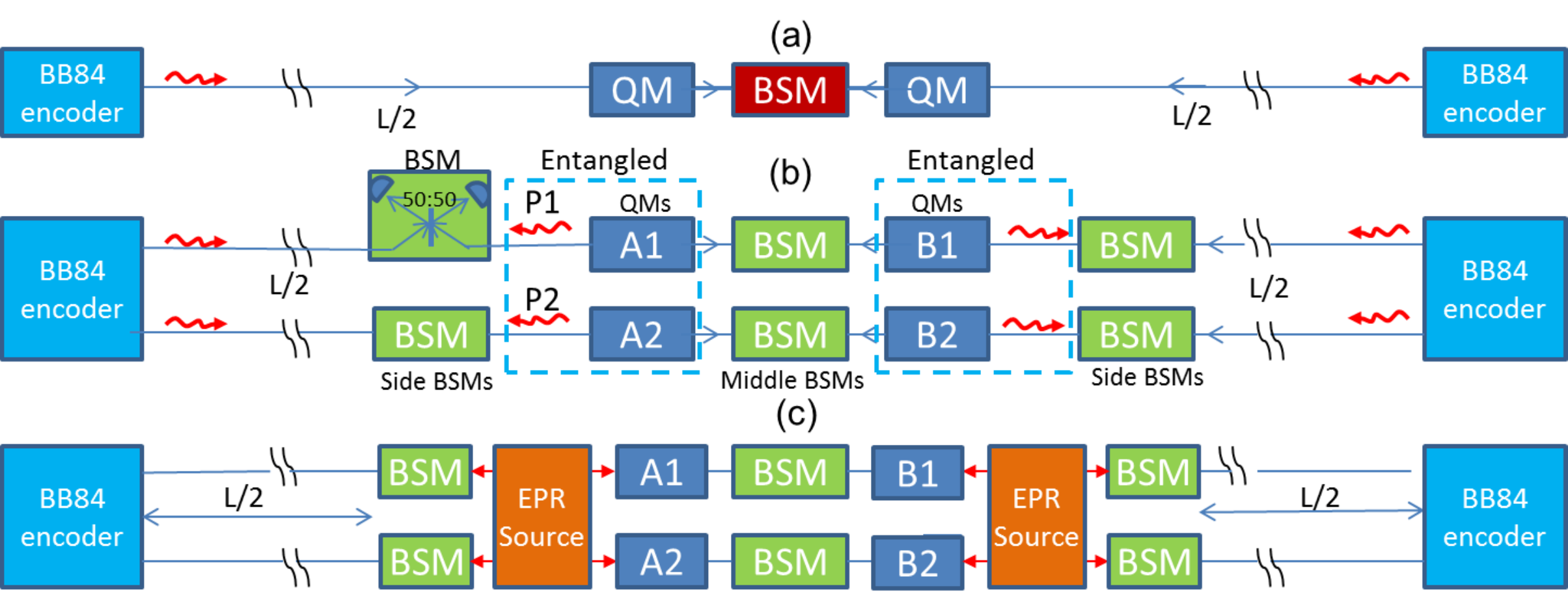} 
\par\end{centering}
\protect\caption{\label{fig:setup_intro}{MA-MDI-QKD schemes with (a) heralding and
(b) non-heralding QMs. In (a), we assume that, using certain
mechanisms, the transmitted photon by the user can be written into
the QM and the memory can herald its successful loading \cite{panayi}. In (b),
the dual-rail configuration for ensemble-based QMs is shown. Here,
in each round, one entangles QMs $A_{1}$ and $A_{2}$, and, similarly,
$B_{1}$ and $B_{2}$, with two optical modes in the vacuum or single-photon
state. At the transmitters, users encode their bits using phase encoded
BB84 as explained in \cite{MDIQKD_finite_PhysRevA2012}. The BSM is
performed using two single-photon detectors and a 50:50 beam splitter
on each rail; see the BSM box for memory A1. All other BSM boxes in
(b) and (c) are the same. A click on only one detector would herald
success for the corresponding BSM. Once both BSMs on one side are
successful, we assume that the user's state has been teleported to
the corresponding QMs. One then continues with loading the other two
QMs, and, once done, they will proceed to perform the middle BSMs.
(c) MA-MDI-QKD with EPR sources. At each round, one generates an entangled
state in the form $|\psi_{{\rm entg}}\rangle_{AP}$, use half of it
to do the side BSM, and, if successful, attempt to store the other
half in the QMs. Note that the dual-rail configuration in (b)
and (c) is for illustration purposes only. In practice, one can use
the equivalent single-rail time-bin encoding techniques.} }
\end{figure}

MA-MDI-QKD is a simple, but effective, extension of MDI-QKD, which
inherits its resilience against detector attacks, and enhances its
rate scaling. In MA-MDI-QKD, the photons transmitted by the users
are stored each in a QM before the entanglement-swapping Bell-state
measurement (BSM) in the middle; see Fig.~\ref{fig:setup_intro}(a).
This setup resembles a quantum repeater link with nesting level one,
because of which rate enhancement follows, but without any QMs at
the end users. The users instead need to have a BB84 encoder, which not only makes the implementation of the system easier,
but also has an additional operational advantage: Now, the repetition
rate of the protocol is not determined by the distance, or the transmission delay, 
between the end users. Instead, one can in-principle run the protocol
as fast as our QMs and optical sources allow without the need to wait for classical signals
to acknowledge the success of entanglement distribution. If one employs
QMs that feature short light-matter interaction times, then one may improve
the total key generation rate per unit of time. One still, however,
requires that the storage of the photon in the QM to be {\em heralding}.
Direct heralding mechanisms for writing photons into QMs are often
slow, because of which the authors in Ref. \cite{panayi} suggested
to use the teleportation idea. That is, by first entangling a photon
with the QM, and performing a side BSM on this photon and the photon
sent by the user, one can indirectly herald the transfer of the user's
state to the corresponding QM.

One of the first investigations \cite{IEEE1} of the above technique utilized atomic-ensemble based QMs in conjunction with a heralding scheme based on off-resonant Raman interactions \cite{DLCZ}. By using such a scheme \cite{DLCZ} for interaction between weak pump signals
and atomic ensembles, one can generate states with dominant terms in
the form (neglecting normalization factors throughout this section)
$|0\rangle_{P}|0\rangle_{A}+\sqrt{p_{c}}|1\rangle_{P}|1\rangle_{A}$,
where $|0\rangle_{P}$ and $|1\rangle_{P}$ are, respectively, vacuum
and single-photon states, $|0\rangle_{A}$ represents an ensemble
with all atoms in their ground states, and $|1\rangle_{A}$ is an
ensemble with only one atom, randomly, in a meta-stable excited state,
while the rest are in the ground state. Using two of such states,
see Fig.~\ref{fig:setup_intro}(b), plus post-selection succeeding
with a rate proportional to $p_{c}$, one can then end up with an entangled
state between two ensembles $A_{1}$ and $A_{2}$, and their corresponding
photonic modes $P_{1}$ and $P_{2}$, in the form of $|\psi_{{\rm entg}}\rangle_{AP}=|0\rangle_{P_{1}}|0\rangle_{A_{1}}|1\rangle_{P_{2}}|1\rangle_{A_{2}}+|1\rangle_{P_{1}}|1\rangle_{A_{1}}|0\rangle_{P_{2}}|0\rangle_{A_{2}}$,
provided that $p_{c}$, the excitation probability, is much lower
than one. The setup in Fig.~\ref{fig:setup_intro}(b) was investigated
in \cite{IEEE1} and it turned out that primarily the $|1\rangle_{P_{1}}|1\rangle_{A_{1}}|1\rangle_{P_{2}}|1\rangle_{A_{2}}$
state, which would be generated with probability $p_{c}^{2}$, could
result in such an amount of error that would prevent this system from
outperforming no-QM systems. We refer to this issue by the two excited
QM (TEQM) problem. Note that reducing $p_{c}$ would also reduce the
success rate of the post-selection mechanism, and, on balance, would
not result in an overall rate advantage.

There are several solutions to the TEQM problem. First, one may consider only quasi-single-atom QMs,
such as nitrogen-vacancy (NV) centers in diamond, as proposed in Ref.
\cite{MA_NV_centers}. In order to obtain a significant improvement in the key rate however, the NV centers must be embedded into microcavities \cite{MA_NV_centers}. While it is shown that the required cavity cooperativity is not necessarily high, their entangling protocol requires an appropriate SPS to entangle a photon with the electron spin of the NV center \cite{MA_NV_centers}, a combination of which has yet to be demonstrated. Another remedy to TEQM, proposed in Ref. \cite{IEEE1}, is to
use nearly ideal entangled-photon (EPR) sources for creating the initial
QM-photon entanglement; see Fig.~\ref{fig:setup_intro}(c). The idea
is that if one has an EPR source that ideally generates only one pair
of photons per trigger, and that one can efficiently store one of these
photons into a QM without concerns of multiple-excitation
issues. In Ref. \cite{IEEE1}, the authors show that conventional
EPR sources relying on parametric down-conversion would not solve
the problem, but suggest to instead use quantum-dot based EPR sources,
of which have been shown to have very low second-order coherence properties
\cite{QDot_entg_low_g2:NatPhot2014}. Similarly, this solution would benefit quantum repeater implementations \cite{spdc_repeater}. The other benefit of the EPR-based
approach is that one only needs to write into QMs if the corresponding
side-BSM is successful. We refer to this technique as ``delayed writing'',
which further reduces the requirements on the access times of QMs
as they do not need to be initialized in every round.

Our proposed solutions consider SPSs as a replacement for EPR sources for implementing the above ideas. SPSs are at a more advanced stage of development than EPR sources, which opens up the possibility of a proof-of-principle experiment to be accomplished in the short term. For instance, SPSs relying on quantum dot structures \cite{SomaschiN.2016} can offer high-rate and low-noise performance that is suitable for the systems we propose here.

Specifically, we propose two setups. The first, of which resembles a noiseless linear amplifier (NLA) \cite{NLA}, involves an entangling procedure that is based on the method described in Refs. \cite{SPS,LoPiparo:2013}. Simply, the user's photons are passed through a NLA before storing them into the QMs. For this setup, we optimize over the NLA parameters to maximize the key rate, finding that, while the TEQM issue is resolved, the rate scaling does not improve.  Our second, improved, solution, consists of a ``quasi-EPR'' source relying on two SPSs. This setup provides the required entanglement after post-selection (via the side BSMs), solves the TEQM issue, improves the rate, and is compatible with some non-ideal SPSs \cite{q_dots_72}.


The key rate of our system not only depends on the entangling procedure
but also on the characteristics of the QMs that are employed \cite{IEEE1}. Thus, we calculate the secret key rate of the proposed MA-MDI-QKD protocols considering different types of ensemble-based QMs. The latter may differ in coherence time, efficiency, bandwidth and access time, reading and writing procedures, and operating wavelength for example. In particular,
we consider a selection of state-of-the-art memories based on warm vapors at room temperature
\cite{cavity_Raman_2016,Wal_mem,mag_shield,ORCA}, cold ensembles
of rubidium atoms \cite{cold1,cold2,cold3}, and cryogenically-cooled rare-earth-ion-doped
crystals (REICs) \cite{151Eu:Y2SiO5,153Eu:Y2SiO5cavity,Pr:Y2SiO5cavity,Pr:Y2SiO5,Pr:Y2SiO5non_classical}.
In the latter case, we utilize the property of ensemble memories to store multiple excitations (each in distinguishable modes) to our advantage, in that we account for the possibility of spectral multi-mode storage \cite{AFC_multi}.

We consider all major sources of errors in each MA-MDI-QKD setup,
such as, channel loss, efficiency and background noise due to photodetection
and frequency conversion, as well as coherence time, and writing-reading
efficiencies of QMs. Based on our calculations, we find existing and
near-future candidates of MA-MDI-QKD systems that offer better performance
than existing QKD links.

The paper is structured as follows. In Sec. II we describe our two
proposed setups. In Sec. III, we study the performance of these setups
by calculating their secret key rates. In Sec. IV, we present our
numerical results by comparing the key rate with the fundamental rate
bounds for the distribution of secure keys over a lossy channel found
in Ref. \cite{Pirandola}. We also determine the secret key rate of the
quasi-EPR-based setup for different types of ensemble-based QMs and
we compare the rate with that of the no-memory systems. In Sec. V,
we draw our conclusions.

\section{System description}

We describe our two SPS-based MA-MDI-QKD setups in this section.
The setups we present here both use the EPR source structure in Fig.~\ref{fig:setup_intro}(c)
except that, instead of the actual EPR source, we use SPS-based modules
that have similar functionality. 
We run the protocol with a repetition rate $R_{S}=1/T$, where $T$
is the repetition period, that is mainly specified by the SPS. In both cases we assume that the delayed writing
procedure is used. That is, one attempts to write the photons into
the QMs $A_{1}$ and $A_{2}$ only if both corresponding side BSMs
are successful, and do similarly for $B_{1}$ and $B_{2}$. The delay
required for this step can be on the order of nano seconds, corresponding
to the measurement time at the BSMs \cite{Yuan:Selfdif:2007}, and
should not incur much additional loss or complexity. The benefit is that the potentially time-consuming initialization of the QMs shall only be done once the memories have been loaded and read instead of in every round. The loading/reading step occurs at a much lower rate especially
at long distances. In the following, we first explain our NLA- and
quasi-EPR-based setups, and then give a precise description of all
components used in these systems.




\subsection{NLA-based MA-MDI-QKD}

\begin{figure}
\begin{centering}
\includegraphics[scale=0.6]{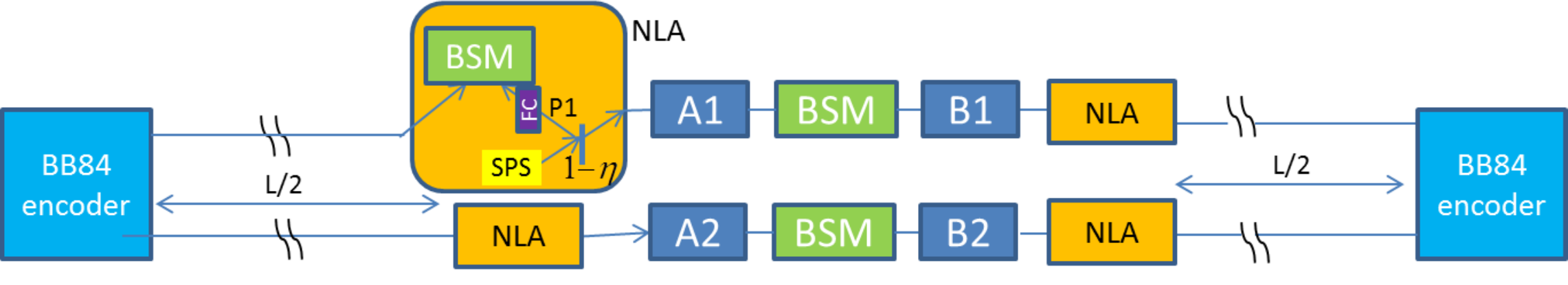} 
\par\end{centering}
\protect\caption{\label{fig:setup_NLA}{The NLA-based MA-MDI-QKD. Users' photons will
be effectively amplified before being stored in the QMs. A successful
loading would be declared if the two NLAs corresponding to each user
are both successful, that is, their corresponding BSM modules have
a single click. We use the same BSM modules as in Fig.~\ref{fig:setup_intro}(b).
The FC box represents the frequency converter.} }
\end{figure}

The key requirement of the setups of Figs.~\ref{fig:setup_intro}(b)
and (c) is to generate entanglement between photons and QMs. Our aim is to achieve the same objective by using
SPSs. A solution that may by envisioned utilizes entanglement
distribution techniques that rely on SPSs. Not surprisingly, there is a class
of probabilistic quantum repeaters that have such a property. In the
scheme proposed in Ref. \cite{SPS}, the authors use a SPS and an
imbalanced beam splitter to create spin-photon entanglement. They interfere two such photonic modes at a BSM to entangle the corresponding QMs. In Fig.~\ref{fig:setup_NLA}, we have used
a similar idea to create our desired entangled state in the form $|\psi_{{\rm entg}}\rangle_{AP}$,
for $A_{1}$ and $A_{2}$ and their corresponding photonic modes $P_{1}$
and $P_{2}$. The same can be done for $B_{1}$ and $B_{2}$ in Fig.~\ref{fig:setup_NLA}.
Here, we use a beam splitter with reflectivity $\eta$ to split a
single photon into two paths: one would be stored into a QM, and the
other interferes at a BSM with the signal sent by the user. This
structure, as shown for memory $A_{1}$ in Fig.~\ref{fig:setup_NLA}, then resembles a NLA module based on quantum scissors \cite{NLA}. We consider the QMs to be loaded with the user's transmitted state (within a known rotation) if both NLAs on one side are successful, meaning that their BSM module generates exactly one click, 

As shown below, the above NLA structure can provide us with the required entanglement. Suppose the writing efficiency into QMs
is unity, and, without loss of generality, let us focus on QMs
$A_{1}$ and $A_{2}$. Assuming ideal on-demand SPSs, the joint state
of the QMs and their corresponding optical modes $P_{1}$ and $P_{2}$
is given by 
\begin{equation}
\begin{array}{c}
|\psi{\rangle}_{AP}=\sqrt{\eta\left(1-\eta\right)}\left(|10\rangle_{P_{1}P_{2}}|01\rangle_{A_{1}A_{2}}+|01\rangle_{P_{1}P_{2}}|10\rangle_{A_{1}A_{2}}\right)\\
+\eta|11\rangle_{P_{1}P_{2}}|00\rangle_{A_{1}A_{2}}+(1-\eta)|00\rangle_{P_{1}P_{2}}|11\rangle_{A_{1}A_{2}},
\end{array}\label{eq:NLA_dm}
\end{equation}
where the first term, in brackets, is the desired entangled state.
After the postselection by the two BSMs, which requires exactly one
click in each module, the last term in \eqref{eq:NLA_dm} would be
ideally removed. This last term is what could cause the TEQM problem.
Therefore, this scheme resolves the TEQM issue. There is, however,
a remaining term in the form $|11\rangle_{P_{1}P_{2}}|00\rangle_{A_{1}A_{2}}$,
which is unwanted but can result in successful BSMs with a probability
proportional to $\eta^{2}$, {\em whether or not} the transmitted
photons have survived the path loss. That is, because of one background
photon in each leg, in the asymptotic limit, when the distance $L$ is
large, the success rate of the side BSMs is nonzero. Let us give
a name to this issue and call it the ``two loss-independent click'' (TLIC)
problem. We will show in Sec.~\ref{sec:Secret-key-rate} how this
problem prevents us from getting any rate advantage over no-QM
setups. The scheme of Ref. \cite{IEEE1} as shown in Fig.~\ref{fig:setup_intro}(b)
also suffers from the TLIC issue. Note again that reducing $\eta$
alone may not solve the problem, as our desired term occurs with a
probability proportional to $\eta$. In principle, dark counts could
also cause the TLIC problem, but, we may ignore it for now
if it is small in comparison with other sources of background photons.
We comment on the effect of dark counts later in this section and
fully account for it in our key rate analysis. Next, we examine another
solution that resolves the TLIC problem as well as the TEQM one.

\subsection{MA-QKD with Quasi-EPR Sources}

\begin{figure}
\begin{centering}
\includegraphics[scale=0.6]{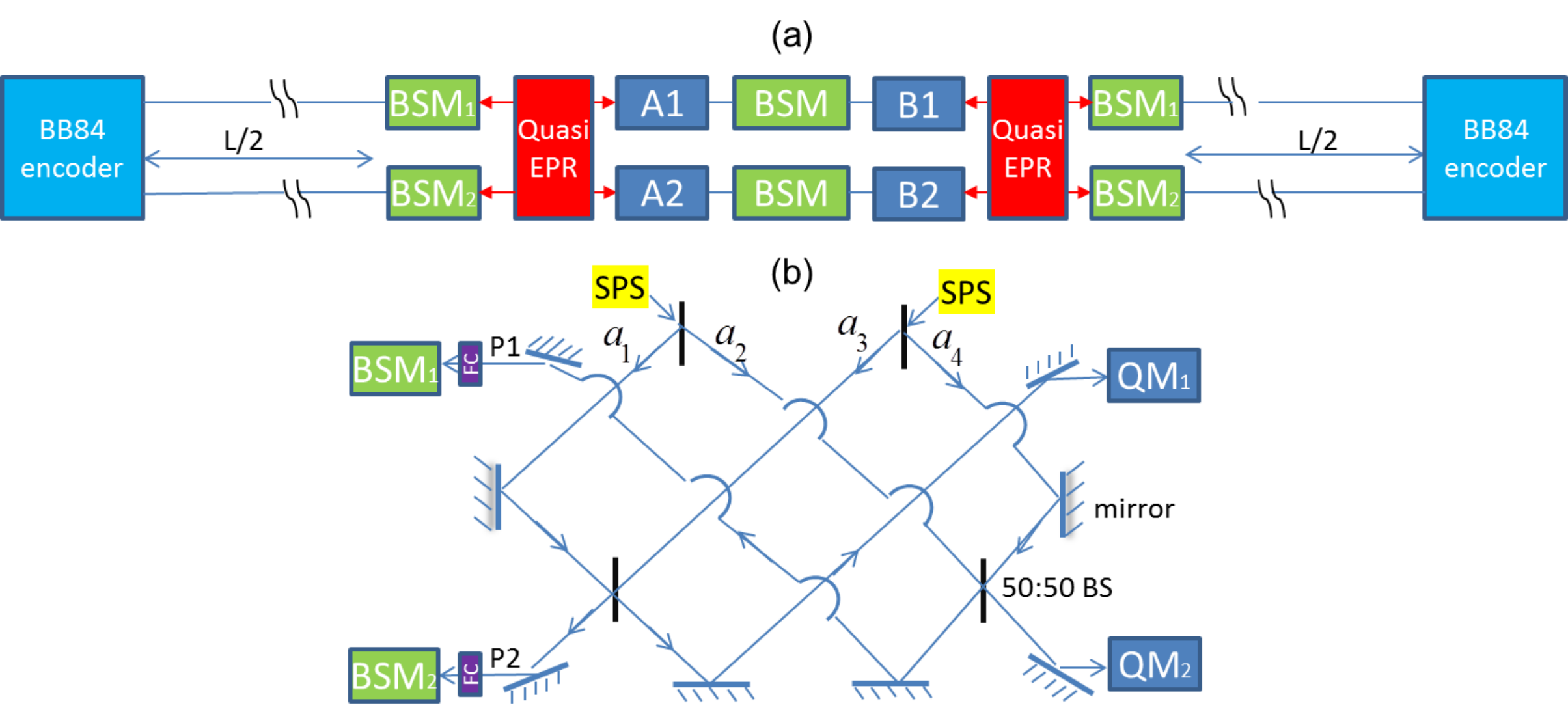} 
\par\end{centering}
\protect\caption{\label{fig:setup_quasi}{(a) MA-MDI-QKD with quasi-EPR sources. The
quasi-EPR source, shown in (b), relies on two SPSs and a network of
50:50 beam splitters and mirrors that interfere the two photons via
different paths. If both side BSMs are successful, the post-selected
state resembles an entangled state in the desired form. }}
\end{figure}

Figure \ref{fig:setup_quasi}(a) shows the MA-MDI-QKD setup with quasi-EPR
sources for entangled photons. Our proposed quasi-EPR module is shown
in Fig.~\ref{fig:setup_quasi}(b), of which may be built using integrated
optics. It produces the desired entangled states, by interfering
two single photons at different balanced beam splitters. It also generates
additional spurious terms, which we aim to select out after successful
side BSMs. Analyzing the circuit in Fig.~\ref{fig:setup_quasi}(b),
and using ideal $A_{1}$ and $A_{2}$ memories, the joint state
of $A$ and $P$ modes can be written as follows 
\begin{equation}
\begin{array}{c}
|\psi\rangle_{AP}=1/2\left(|10\rangle_{P_{1}P_{2}}|10\rangle_{A_{1}A_{2}}-|01\rangle_{P_{1}P_{2}}|01\rangle_{{A_{1}A_{2}}}\right)\\
+\frac{1}{2\sqrt{2}}\left(|20\rangle_{P_{1}P_{2}}+|02\rangle_{P_{1}P_{2}}\right)|00\rangle_{{A_{1}A_{2}}}\\
+\frac{1}{2\sqrt{2}}\left(|20\rangle_{{A_{1}A_{2}}}+|02\rangle_{{A_{1}A_{2}}}\right)|00\rangle_{P_{1}P_{2}},
\end{array}\label{eq:EPR_dm}
\end{equation}
where, again, the first term, in brackets, on the right-hand side,
represents the desired entangled state. The last term represents the
no-photon term, hence, unless for negligible dark count effects, cannot
result in successful side BSMs, and it would be selected out. The
term in the middle could result in successful BSMs, provided that
the user's photon survives the path loss and/or because of dark
counts. But, then, the QMs are both in their ground states, and
except for a probability proportional to the dark count rate, they
will not produce successful results at the middle BSM, and will be
selected out at that stage. 

Specifically, by proper use of the quantum interference effect, in Fig.~\ref{fig:setup_quasi}(b),
we have managed to group the unwanted states into terms that ensures both photons
appear at the same output port. This creates only {\em one} background-induced click, making it easier to remove them by postselection. In the case of ideal SPSs, the above solution does resolve both the TEQM
and TLIC problems. Even in the case of the second term, in order to
get two successful side BSMs, one needs to have a user's photon arriving
at the receiver, whose probability goes to zero at large distances.
All of the previous discussion is based on the assumption that dark counts are negligible. The situation
would be different if we have non-ideal SPSs with non-zero probabilities
for emitting more than one photon, or when we have substantial dark
count or background noise. We will consider theses scenarios later
in our paper.


We made some idealistic assumptions in explaining how our proposed
entanglement generation processes work. In the next section, we properly
model major non-idealities in the system from which a realistic account
of the key rate performance can be obtained.

\subsection{Device modeling}

\label{Sec:Device} We model different components of our system as
follows.

\textbf{BB84 encoders:} We use phase encoding in the dual rail setup,
or, equivalently, and if allowed by the QM setups, time-bin encoding in a single-rail setup, in bases
$Z$ and $X$ \cite{MDIQKD_finite_PhysRevA2012}. We assume that efficient
QKD protocols are in use \cite{Lo:EffBB84:2005}, where basis $Z$ is chosen most often. We also assume that both users employ ideal
SPSs for their BB84 encoding. In principle, they can use the decoy-state
version of BB84, but, for the sake of our comparison, it would be
sufficient to assume that both QM-assisted and no-QM systems use single
photons to encode their bits. The multi-photon terms in a decoy-state
protocol can be characterized by statistical analysis and they will
not impose a change in the rate scaling \cite{MDIQKD_finite_PhysRevA2012}.
The pulse duration is denoted by $\tau_{p}$, and it is assumed to
be equal to $T$ in our numerical analysis.

\textbf{Channel:} We denote the total channel length by $L$, and
its attenuation length by $L_{{\rm att}}$. That is, the total channel
transmissivity will be given by $\exp(-L/L_{{\rm att}})$. We assume
that the channel does not impose any phase or polarization distortions.
In practice, such effects can be compensated by classical-feedback mechanisms. The error
in such compensating mechanisms can then be analytically modeled via
misalignment parameters. In this work, we neglect such errors as they
are not major error bearing components of our system, and they are
common for both QM-assisted and no-QM systems. One can use the methods
proposed in Refs. \cite{panayi,IEEE1} to account for such imperfections.

\textbf{Single-photon detectors (SPDs):} All our employed SPDs are
assumed to be non-resolving detectors with efficiency $\eta_{D}$.
The dark count rate is denoted by $\gamma_{dc}$, which results in
a dark count probability $d_{c}=\gamma_{dc}\tau_{p}$ per pulse. Here
we assume that photodetectors are gated with an opening time that is identical
to the pulse duration. The time that it takes to detect
a photon and prepare the detector for next measurement is denoted by $\tau_{M}$. Using self-difference techniques \cite{Yuan:Selfdif:2007}, $\tau_{M}$ can be on the order of nanoseconds.

\textbf{Quantum memory:} We consider several characteristics of QMs
pertinent to our setups. The writing efficiency into QMs, i.e., the
probability of successfully transferring the qubit-state encoded into a single photon
to the QM, is denoted by $\eta_{w}$. The probability of successfully
reading the QM, i.e., transferring the qubit-state encoded into a QM (back) onto a single photon
is denoted by $\eta_{r}$. The latter will be affected by amplitude
decay with time constant $T_{r}$. The reading efficiency at time
$t$ after the loading is then assumed to be given by $\eta_{r}=\eta_{r0}\exp\left(-t/T_{r}\right)$
\cite{panayi,IEEE1}, where $\eta_{r0}$ is the reading efficiency
right after the loading. The exponential decay is not necessarily
the case for all memories studied in this work. For instance, the decay is Gaussian for AFC-based QMs that do not compensate for dephasing induced by ground-level inhomogeneous broadening. In the regime of interest, where the relevant system time parameters are shorter than $T_{r}$, the exponential decay assumption will then be a pessimistic one for such QMs and it would not alter the overall conclusions made in our work. We also denote the required time to initialize the memory by $\tau_{{\rm init}}$ and the time needed to interact with single photons by $\tau_{{\rm int}}$.

\textbf{Single-photon source:} We assume that the SPSs used in the
middle site of Figs.~\ref{fig:setup_NLA} and \ref{fig:setup_quasi}
are identical but probabilistic. That is, upon trigger, there is a
likelihood $\eta_{{\rm SPS}}$ that they generate the following normalized
state 
\begin{equation}
\rho_{{\rm SPS}}=p_{1}|1\rangle\langle1|+p_{2}|2\rangle\langle2|,
\end{equation}
where $|2\rangle$ is the two-photon state, and $p_{1}\eta_{{\rm SPS}}$
and $p_{2}\eta_{{\rm SPS}}$ are, respectively, the single-photon
and double-photon probabilities. For most of this paper, we assume
that $p_{2}=0$. We will examine the range of values for $p_{2}$ that are be tolerable for our setups.

\textbf{Frequency Converter:} Given that many QMs do not operate at
the telecom wavelengths, we may need to convert the frequency of some
of the generated photons to match that of the QM or the telecom channel
used. We consider three scenarios: (1) use SPSs that generate
photons at telecom wavelengths. One may then need an upconverter right
before the QMs. The advantage is that the side BSM can be done more
efficiently. On the downside, however, all the errors in the upconversion
will affect the QM as well; (2) generate photons that are matched
to the QM, but we downconvert the photons that enter the side BSM.
Here, the advantage is that one can possibly use a matched SPS, in
terms of the QM bandgap and its bandwidth, to maximize the writing
efficiency, but one will have noisier side-BSMs in this case; and (3), which is similar to (2), but one upconverts the photons sent by the user before
the side BSM. In this work, we adopt the second scenario and assume
that the wavelength and the bandwidth of the SPSs matches that
of QMs. In order to do side-BSMs, one may need to use a down-converter
to match the wavelength of the two interfering photons \cite{Upconversion_1558to662,Upconversion_1550to800,Downconversion_738_to_1550}.
We account for the conversion efficiency of such devices in our analysis.
We also assume that the additional background Raman photons generated
by the down-converter would modify the dark count of the side-BSM
detectors.

In all devices, the sources of inefficiency are modeled by fictitious
beam splitters with proper transmissivities.

\section{Key rate analysis\label{sec:Secret-key-rate}}

In this section, we find the secret key generation rate for our proposed
schemes shown in Figs.~\ref{fig:setup_NLA} and \ref{fig:setup_quasi}.
We assume that there is no eavesdropping and we are only affected
by device imperfections of the system as modeled in Sec.~\ref{Sec:Device}.
For convenience, we assume that both setups are symmetric. Under these
conditions, in the infinite-key setting, the secret key generation
rate in the setups of Figs.~\ref{fig:setup_NLA} and \ref{fig:setup_quasi}
is lower bounded by
\begin{equation}
R_{{\rm QM}}=R_{S}Y_{11}^{{\rm QM}}\left[1-h(e_{11;X}^{{\rm QM}})-fh(e_{11;Z}^{{\rm QM}})\right],\label{eq:key_rate}
\end{equation}
where $e_{11;X}^{{\rm QM}}$ and $e_{11;Z}^{{\rm QM}}$, respectively,
represent the quantum bit error rate (QBER) between Alice and Bob
in the $X$ and $Z$ basis, and $R_{S}Y_{11}^{{\rm QM}}$ is the rate
at which one generates raw key bits; the index 11 means that single
photons are used at BB84 encoders; $f$ denotes the inefficiency of
the error correction scheme, and $h(q)=-q\log_{2}(q)-(1-q)\log_{2}(1-q)$
is the Shannon binary entropy function \cite{key_rate_bound,panayi}.


We use the techniques of Refs. \cite{panayi,IEEE1} to calculate
the above terms in the scenarios of interest. Without fully repeating
the detailed calculations, here we just highlight the key steps in
the derivation that are important in our understanding of the key-rate
behavior of setups in Figs.~\ref{fig:setup_NLA} and \ref{fig:setup_quasi}.
The key idea behind calculating $Y_{11}^{{\rm QM}}$ is to decompose the
problem into two parts: (1) how often one loads the QMs on both
sides, and (2) once loaded, how often the middle BSMs succeeds. Let
us denote the former by $P_{{\rm SBSM}}$ and the latter by $P_{{\rm MBSM}}$, to give
\begin{equation}
Y_{11}^{{\rm QM}}=P_{{\rm SBSM}}P_{{\rm MBSM}}.\label{eq:Y_QM}
\end{equation}
Here, $P_{{\rm SBSM}}$ partly depends on the probability to obtain two successful side-BSMs on one side, and partly on memory reading and writing times. Once both QMs are loaded, one has to spend
a time equivalent to $\tau_{r}=\tau_{{\rm int}}+\tau_{M}+\tau_{{\rm init}}$
to obtain a measurement outcome for the middle BSM, and prepare the QMs for the next round
\cite{MA_NV_centers}. Accounting for $\tau_{w}=\tau_{{\rm int}}+\tau_{M}$ to write into the QM, there is a minimum time of $\tau_{w}+\tau_{r}$ to get one raw key bit. The inverse of this parameter then sets a bound on the maximum key rate achievable from our delayed-writing schemes. At long distances, however, the challenge of ensuring both sides to be loaded would take precedent, hence we would expect that $P_{{\rm SBSM}}\approx\frac{2}{3}\Pr(\mbox{Successful side-BSMs on one side})$
\cite{panayi}. As for $P_{{\rm MBSM}}$, the difficult part is to account for the decay of the QMs that may be loaded earlier. This requires us to average over the statistics of loading as has been detailed in \cite{panayi,IEEE1}. The same averaging is required in the calculation of $e_{11;X}^{{\rm QM}}$ and $e_{11;Z}^{{\rm QM}}$. Note that when $T_{r}$ is sufficiently large, we can ignore the averaging, and we have $P_{{\rm MBSM}}\approx\Pr(\mbox{two successful middle BSMs})$.

All above terms are found by calculating the relevant output density matrices in the setups of interest. We analytically obtain the pre-measurement state of the system by applying a series of transformations on the input density matrix considering channel transit, the entangling circuits, and the BSM modules. After applying relevant measurement operators, one can then find the post-measurement states and the relevant probability terms. In our case, this has been implemented using a generic Maple code developed for such setups.

Next, we examine the key rate scaling of the NLA-based and the quasi-EPR MA-QKD setups.

\subsection{Key rate scaling: NLA-based setup}

\label{Sec:Rate_NLA} In this section, we investigate how the secret
key rate of the scheme shown in Fig.~\ref{fig:setup_NLA} behaves at long distances. Here we ignore all
inefficiencies except for the channel loss for simplicity. We also assume that $T_{r}$
is sufficiently large. Under these conditions, from Eq.~\eqref{eq:NLA_dm},
there are two major terms that correspond to successful side-BSMs.
The first term in brackets on the right-hand side of Eq.~\eqref{eq:NLA_dm}, corresponding to the desired entangled state, would result in successful side-BSMs provided that the user's photon has survived the path loss. This happens with a probability proportional to $(1-\eta)\eta\exp(-L/2/L_{{\rm att}})$ for which the QMs are left in the desired state. The other term that could result in successful side-BSMs is $|11\rangle_{P_{1}P_{2}}|00\rangle_{A_{1}A_{2}}$,
which succeeds with probability $\eta^{2}$ and would leave the QMs
in their ground state. At long distances, the post-measurement state
of the QMs would be roughly given by 
\begin{equation}
\rho_{A_{1}A_{2}}^{{\rm (PM)}}=\frac{\eta^{2}}{P_{{\rm SBSM}}}|00\rangle_{A_{1}A_{2}}\langle00|+\frac{(1-\eta)\eta e^{-L/2/L_{{\rm att}}}}{P_{{\rm SBSM}}}\rho_{A}^{{\rm (TX)}},\label{A1A2PM}
\end{equation}
where 
\begin{equation}
P_{{\rm SBSM}}\propto\eta^{2}+(1-\eta)\eta e^{-L/2/L_{{\rm att}}}\label{PSBSM}
\end{equation}
and $\rho_{K}^{{\rm (TX)}}$ represents the transmitted state (up
to a known rotation) by user $K=A,B$. Starting with $\rho_{A_{1}A_{2}}^{{\rm (PM)}}\otimes\rho_{B_{1}B_{2}}^{{\rm (PM)}}$,
then, for the middle BSM, one has 
\begin{equation}
P_{{\rm MBSM}}\approx\frac{1}{P_{{\rm SBSM}}^{2}}\left[\eta^{4}d_{c}^{2}+2\eta^{3}(1-\eta)e^{-L/2/L_{{\rm att}}}d_{c}+\eta^{2}(1-\eta)^{2}e^{-L/L_{{\rm att}}}\right].
\end{equation}
In the regime of operation where $\eta\gg(1-\eta)e^{-L/2/L_{{\rm att}}}\gg d_{c}$,
we then obtain 
\begin{equation}
Y_{11}^{{\rm QM}}=P_{{\rm SBSM}}P_{{\rm MBSM}}\propto(1-\eta)^{2}e^{-L/L_{{\rm att}}},
\end{equation}
that is, the key rate scales with the loss in the entire channel as is the case for a conventional no-QM system, and one should not expect any benefit from the NLA-based setup. As mentioned before, the distance-independent
terms in Eqs.~\eqref{A1A2PM} and \eqref{PSBSM} are the root causes
of the TLIC problem. The dependence of both desired and undesired
terms on $\eta$ is another factor that results in such a rate scaling,
even if one ignores the error terms. The condition $\eta\gg(1-\eta)e^{-L/2/L_{{\rm att}}}\gg d_{c}$
represents operating regime of interest when long distances are considered
as we show in Sec.~\ref{sec:Numerical-results}.

\subsection{Key rate scaling: Quasi-EPR setup}

\label{sec:QEPR} Using a similar analysis as in the previous section,
we calculate how the secret key rate scales for the quasi-EPR
setup of Fig.~\ref{fig:setup_quasi}. In this case, for an ideal
SPS with $p_{1}=1$ and $p_{2}=0$ and ignoring dark counts, from Eq.~\eqref{eq:EPR_dm},
one has $P_{{\rm SBSM}}\propto\exp(-L/2/L_{\mathrm{att}})$, as none
of the terms will generate detection events corresponding to both BSMs  in the cases whereby the users' photons are lost. This implies that $\rho_{A_{1}A_{2}}^{{\rm (PM)}}$
is in the form $\alpha|00\rangle_{A_{1}A_{2}}\langle00|+\beta\rho_{A}^{{\rm (TX)}}$
for some constant $\alpha$ and $\beta$, adding up to one. $P_{{\rm MBSM}}$
would then be given by 
\begin{equation}
P_{{\rm MBSM}}\approx\frac{1}{P_{{\rm SBSM}}^{2}}\left[\alpha^{2}d_{c}^{2}+2\alpha\beta e^{-L/2/L_{{\rm att}}}d_{c}+\beta^{2}e^{-L/L_{{\rm att}}}\right],
\end{equation}
which, when $d_{c}\ll e^{-L/2/L_{{\rm att}}}$, results in 
\begin{equation}
Y_{11}^{{\rm QM}}\propto e^{-L/2/L_{\mathrm{att}}}.\label{eq:R_EPR}
\end{equation}
From Eq. \eqref{eq:R_EPR}, we infer that the key rate for the setup
of Fig.~\ref{fig:setup_quasi} scales similarly as a single-node
quantum repeater system. Although we have not yet accounted for the
errors in this rough analysis, the quasi-EPR setup promises to outperform the no-memory schemes. We examine this conjecture in the next section.

The above conclusion relies on the assumption that $p_{2}=0$ and results in $P_{{\rm SBSM}}$ being proportional to the channel loss. If the probability to obtain the two-photon terms of the SPS is non-zero, then the distance-independent terms in $P_{{\rm SBSM}}$ are on the order of $p_{2}$, similar to $\eta^{2}$ in Eq.~\eqref{PSBSM}. Such terms would result in the TLIC problem as before once $p_{2}/p_{1}$
is comparable to $e^{-L/(2L_{{\rm att}})}$. The same holds if $d_{c}$
is on the order of $e^{-L/(2L_{{\rm att}})}$, which could happen
if the frequency converters generate a large background noise. In
the following section, we explore the requirements on the employed
devices in practical setups.

\section{Numerical results \label{sec:Numerical-results}}

In this section we calculate the secret key rate that can be achieved
using the schemes illustrated in Figs.~\ref{fig:setup_NLA} and \ref{fig:setup_quasi}.
Specifically, we first calculate the secret key rate with the assumption
that {\em ideal} QMs, meaning those that feature no limitations
in performance, are employed. We compare the secret key rate per pulse
of both schemes to the maximum rate achievable over a lossy channel,
which we refer to as the PLOB bound \cite{Pirandola}. We find that
the quasi-EPR scheme outperforms the bound, while the NLA scheme,
due to the TLIC problem, fails to surpass the PLOB bound. Next we
calculate the secret key rate, in bits per second, corresponding to
the quasi-EPR scheme in conjunction with experimentally-measured properties
of state-of-the-art warm and cold atomic ensembles as well as solid-state
QMs based on REICs. For comparison we also plot the secret key rate
for a no-memory MDI-QKD implementation driven at 1 GHz repetition
rate; we use the ``no-memory'' label to refer to this system. We
find that, under certain assumptions, some cold atom memories can surpass the no-memory bound
due to their favorable coherence properties.

We also calculate the secret key rate of the quasi-EPR scheme with the assumption that we employ QMs that feature properties with \textit{modest} improvements
over the state-of-the-art memories. We refer to these as ``near-future''
QMs, and find that almost all near-future memories can outperform
the no-memory system. We conclude with a discussion around other possible
sources of imperfection, such as multi-photon events and background
noise, and explore how these impact the quasi-EPR scheme.

\begin{table}
\begin{centering}
\begin{tabular}{|c|c|}
\hline 
Attenuation length, $L_{\mathrm{att}}$  & 17.3 km\tabularnewline
\hline 
Detection efficiency, $\eta_{D}$  & 0.93 at 1550~nm; 0.6 at 800 nm\tabularnewline
\hline 
Dark count rate, $\gamma_{{\rm dc}}$  & 1 cps at 1550~nm; 1000 cps at 800 nm\tabularnewline
\hline 
Error correction inefficiency, $f$  & 1.16\tabularnewline
\hline 
Frequency conversion efficiency  & 0.68\tabularnewline
\hline 
SPS efficiency, $\eta_{{\rm SPS}}$  & 0.72 \tabularnewline
\hline 
\end{tabular}
\par\end{centering}
\protect\caption{\label{tab:List-of-parameters}{List of common parameters and their
nominal values used in our simulations. The channel loss corresponds
to 0.25~dB/km. The detector efficiency and dark count at around 1550 nm
correspond to the superconducting telecom-wavelength detectors reported
in \cite{dark_count}, which can be used at side BSMs. The parameters
at around 800~nm correspond to commercially available silicon SPDs needed
for the middle BSM. This is justified by the fact that the largest wavelength of operation for the QMs we consider is 850 nm, which corresponds to cesium-based QMs, see Section IV B. Similar SPS and FC efficiencies are reported in Refs. \cite{q_dots_72} and
\cite{Downconversion_738_to_1550}, respectively.}}
\end{table}

\subsection{Ideal quantum memories}

\begin{figure}
\begin{centering}
\includegraphics[scale=0.3]{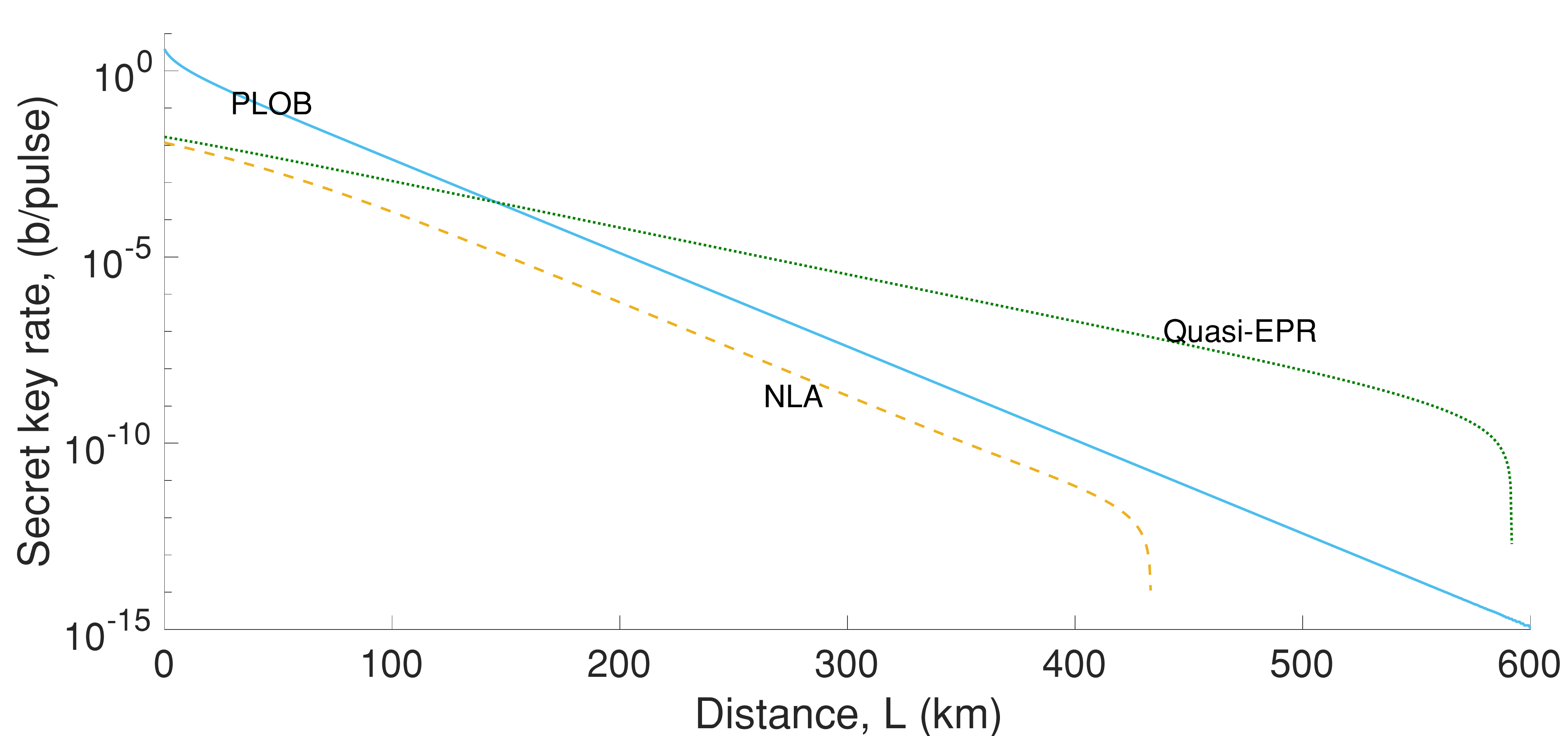} 
\par\end{centering}
\protect\caption{\label{fig:rate_pulse}{Secret key rate per pulse corresponding to
the setups of Figs.~\ref{fig:setup_NLA} and \ref{fig:setup_quasi}.
We have used $\eta=0.2$ for the NLA-based system, which maximizes
the rate. Memories are assumed to be ideal. The only sources of nonideality
are listed in Table~\ref{tab:List-of-parameters} for a detection
gate of 1~ns. We compare the rate with the PLOB bound obtained in
\cite{Pirandola}. }}
\end{figure}

Let us first consider the case of ideal QMs. Specifically, these
memories feature unity reading and writing efficiencies and fidelities,
infinitely long coherence times, unlimited bandwidth, and zero interaction
and initialization times. We calculate the secret key rate for the
NLA and quasi-EPR schemes and that provided by the PLOB bound. The
results are shown in Fig.~\ref{fig:rate_pulse}, where we have used
the values in Table~\ref{tab:List-of-parameters} for the relevant parameters. The NLA-based scheme clearly cannot surpass the PLOB bound,
running below and parallel to it at long distances. Our results validate
the calculations of Sec.~\ref{Sec:Rate_NLA}, which show that, at
long distances, the rate-scaling with distance is the same as a no-repeater
system. If one accounts for imperfections in QMs, the NLA
scheme can only perform worse, which is not promising. This observation can, however, shed some light on the question of whether NLAs can help discrete-variable QKD systems, as compared to the continuous-variable QKD schemes, where, for the latter, some improvement is expected \cite{CVQKD_NLA}. 

Due to its improved rate-versus-distance scaling, the quasi-EPR scheme
can, however, beat the PLOB bound at distances roughly greater than
150~km. Note that this scheme improves the key rate by nearly $\mbox{5}$
orders of magnitude over the PLOB bound at a distance of $700$~km.
Based on this performance, in the following sections, we only focus
on the quasi-EPR scheme for practical and near-future QMs.

\subsection{State-of-the-art quantum memories}

Here we evaluate the performance of the quasi-EPR scheme using a selection
of state-of-the-art ensemble-based memories. There are a variety of
systems that have been utilized for optical QMs; see Ref. \cite{Khabat_review}
for a recent overview. We consider ensemble-based memories due to
their strong light-matter coupling and, in several cases, the
possibility of long coherence times (up to seconds \cite{cold1}) and high bandwidths
(up to several GHz \cite{Wal_mem}). Furthermore, they offer the possibility of multi-mode
storage \cite{Khabat_review,repeaters_review_2011}. By multi-mode we are referring to memories that can simultaneously store more than one qubit during a single storage event by encoding many qubits each into a different mode. This feature has been exploited to enhance secret key generation rates in certain quantum repeater schemes \cite{repeaters_review_2011,AFC_multi,Rate-loss_repeater}. It is important to stress that the definition of multi-mode storage differs from our reference to (the detrimental) storage of multiple excitations. The former involves many excitations, in which each individual excitation occupies a single distinguishable mode (or a pair as required for a a qubit), while the latter concerns many excitations that occupy a single mode and thus each excitation may not be distinguished. 
Motivated by their impressive, and continually-improving, experimental
record, we specifically consider warm vapor (Cs and Rb atomic gas)
and cold atom (Rb atoms in a magneto-optical trap or atomic lattice)
systems \cite{Khabat_review} that rely on the so-called Raman QM
protocol \cite{Raman_memory_theory} as well as cryogenically-cooled
REICs that utilize AFCs \cite{AFC_memory_theory}.

Raman memory schemes \cite{Raman_memory_theory} rely on three energy
levels, usually a $\Lambda$-level system that features long-lived
ground levels. A strong control pulse maps a propagating off-resonant
photon onto the ground level. This is called the ``writing'' step and
at this point the photon is ``stored''. To retrieve the excitation,
a control pulse is applied again, in which the excitation is mapped
back onto a propagating photon. This is referred to as the ``reading''
step. Note that the Raman protocol has been applied to photons that
encode qubits with respect to various degrees of freedom (see Refs.
\cite{Khabat_review,cold2} and references therein). Along with the
convenience of operation at room temperature, warm vapor Raman QMs
feature the possibility to efficiently store GHz-bandwidth photons
with microsecond-long coherence times (with up to 100 $\mu$s being possible
\cite{mag_shield}) \cite{cavity_Raman_2016,Wal_mem,Khabat_review}.
Cold atoms reduce the impact of collisional or motional-induced decoherence,
and, if magnetic-field-insensitive states are used, they offer very
long coherence times reaching hundreds of ms and possibly more \cite{cold1,cold3,Khabat_review}.

In a similar way, on-demand AFC QMs also require a $\Lambda$-level system
except here an optical inhomogeneously-broadened transition is tailored
into a series of narrow absorption lines (the ``comb''), each of which
are detuned from each other by an integer multiple of a fixed detuning
\cite{AFC_memory_theory}. A photon is absorbed by the comb, creating
a delocalized atomic excitation and, using an optical control pulse,
the excitation is reversibly-mapped onto a long-lived spin level.
The photon is emitted due to a quantum interference effect between
each absorption line of the comb. Ensembles of rare-earth-ions are
particularly suited for AFC QMs due to the long coherence times
of both the optical (100s of microseconds \cite{macfarlane1987,thiel_review})
and spin (up to milliseconds \cite{macfarlane1987,thiel_review} or even seconds
\cite{6hT}) transitions in conjunction with level structures that
allow for efficient AFCs over $\sim$MHz bandwidths \cite{Khabat_review,macfarlane1987,thiel_review}.

In the following, we study the performance of certain representatives
from each group of memories. In this subsection and next, we focus
mainly on the memory characteristics and neglect two-photon emissions
from the source (i.e., $p_{2}=0$), or other issues that may arise
in the photonic part of the system. We address the latter issues in
Sec.~\ref{Sec:additional}. We also assume that memories feature
no additional noise for the purpose of our simulations except for
the decoherence effect and coupling issues already accounted for.
This assumption is supported by several recent rare-earth AFC \cite{151Eu:Y2SiO5,Pr:Y2SiO5non_classical},
as well as cold and warm Raman experiments \cite{cold2,Wal_mem,ORCA}
that have shown storage of non-classical light. We have ensured that the repetition rate of each QM does not exceed
the corresponding memory bandwidth. Furthermore, the choice of $\tau_{p}=T$ would minimize any inefficiency due to bandwidth mismatch between the source and the QM. In practice, one may need to choose $\tau_{p}$ to be shorter than $T$, in which case its effect on the coupling efficiency must be considered. For all memories considered, we also assume that $\tau_{{\rm init}}=0$ given that these QMs would ideally
go back to the desired initial state after being read out. In practice,
memory re-initialization may be occasionally needed 
to avoid the spread of error. We assume that the frequency at which
the initialization is needed is sufficiently low that it would not
affect our key rate analysis. 

\emph{Warm vapor:} We consider the Raman memory demonstrations of Refs. \cite{cavity_Raman_2016,Wal_mem,mag_shield,ORCA} for our calculations.
Each of the experiments use Cs vapor, except for Ref. \cite{mag_shield}
which uses $^{85}$Rb, and feature memories of varying performance.
See Table \ref{tab:warm-memories} for a list of relevant memory properties.
Specifically, the QM demonstrated in Ref. \cite{Wal_mem} exhibits
a reasonable combination of efficiency and coherence time as well
as low noise, while Ref. \cite{cavity_Raman_2016} uses an anti-resonance
of a Fabry-Perot cavity to suppress four-wave-mixing-induced noise
that is present in Ref. \cite{Wal_mem}. The limitations of coherence
time in these demonstrations are largely due to imperfect magnetic
shielding, allowing magnetic-field-induced dephasing. The experiment
of Ref. \cite{mag_shield} employs exceptional magnetic shielding,
but does not feature storage of non-classical light. Finally, Ref.
\cite{ORCA} uses a ladder energy-level system to achieve storage
in an excited level, which opens the possibility of storage of light pulses of less than $\sim$100 ps duration. The storage time is, however, restricted to $\sim$100 ns.

\begin{figure}
\begin{centering}
\includegraphics[scale=0.3]{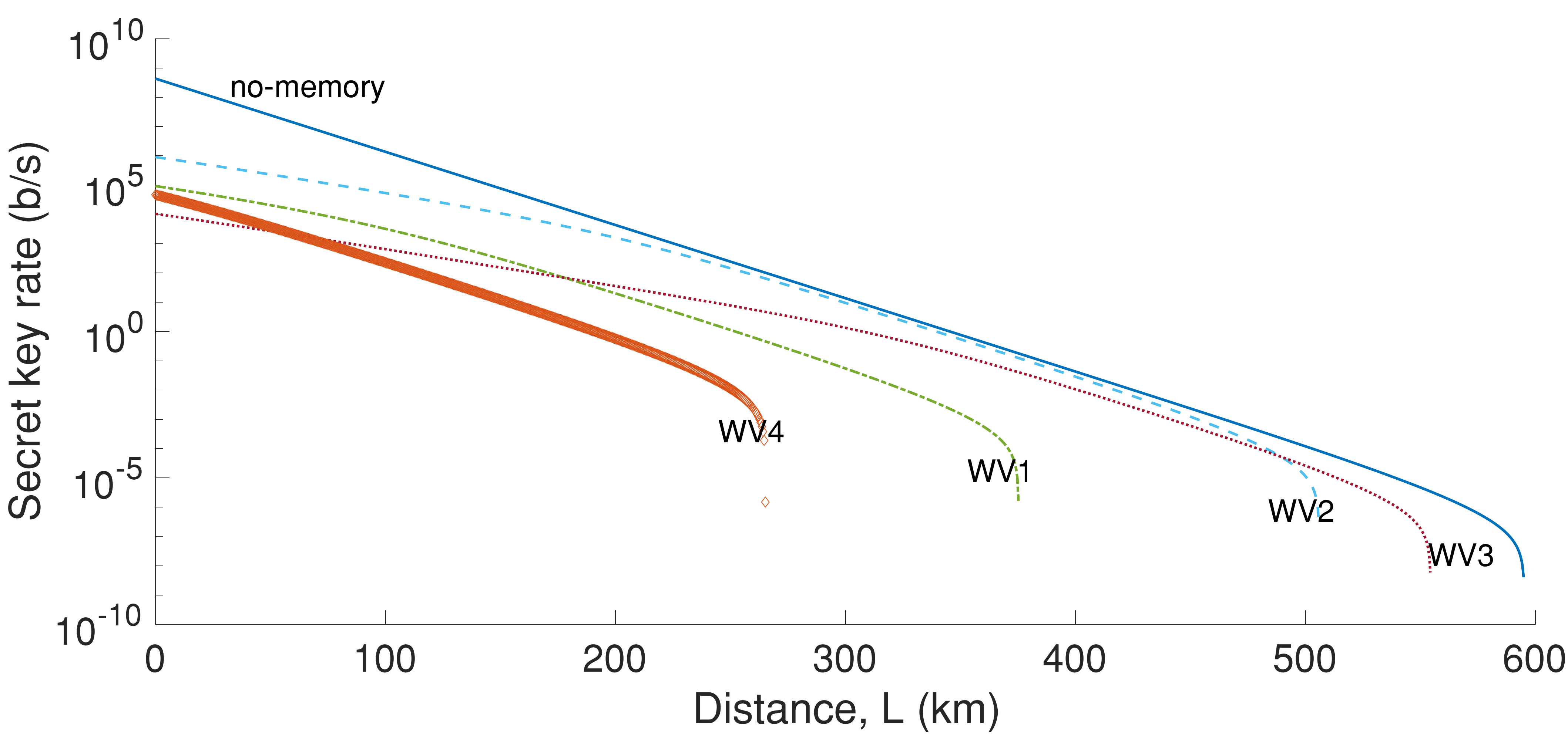} 
\par\end{centering}
\protect\caption{\label{fig:warm-memories}{Secret key rate for the setups of Fig
\ref{fig:setup_quasi} using the parameters of Table~\ref{tab:List-of-parameters}
and the warm vapor memories featured in Table~\ref{tab:warm-memories}.
}}
\end{figure}

Figure \ref{fig:warm-memories} shows the secret key rate of the quasi-EPR
scheme using the memories listed in Table~\ref{tab:warm-memories}
as compared to the no-memory case. It can be seen that none of the
considered QMs can surpass the no-memory curve. Nonetheless, the QM
of Ref.~\cite{Wal_mem} allows the rate to become very close to that of no-memory case, and could surpass the no-memory curve if the QM coherence time was a bit longer or its coupling efficiency was a bit higher. Because of insufficient coherence time, the slope of the curve corresponding to memory WV2 starts changing around 200~km of distance. The lower slope corresponds
to rate scaling with $\exp[-L/(2L_{{\rm att}})]$, whereas the higher
slope corresponds to $\exp[-L/(L_{{\rm att}})]$ scaling, similar
to the no-memory case. The change in slope happens later for WV3, which has the highest coherence time, and much
earlier for the other two QMs. In the case of WV4, the
coherence time is so low that the entire curve is parallel to that
of the no-memory curve, indicating the same rate-versus-distance scaling. In Section C, we show that the no-memory bound may be overcome with some minor improvement in these QMs.

\begin{table}
\begin{centering}
{\scriptsize{}{}}%
\begin{tabular}{|c|c|c|c|c|}
\hline 
\textcolor{black}{\scriptsize{}{}{}}{\scriptsize{}{} }  & {\scriptsize{}{}WV1 \cite{cavity_Raman_2016} }  & {\scriptsize{}{}WV2 \cite{Wal_mem} }  & {\scriptsize{}{}WV3 \cite{mag_shield} }  & {\scriptsize{}{}WV4 \cite{ORCA} }\tabularnewline
\hline 
\textcolor{black}{\scriptsize{}{}Efficiency, $\eta_{w}\eta_{r0}$}{\scriptsize{}{}
}  & {\scriptsize{}{}{{}0.1} }  & {\scriptsize{}{}{{}0.3} }  & {\scriptsize{}{}{{}0.05} }  & {\scriptsize{}{}{{}0.15}}\tabularnewline
\hline 
{\scriptsize{}{}
}\textcolor{black}{\scriptsize{}{}{}Coherence time, $T_{r}$}{\scriptsize{}{}
}  & \textcolor{black}{\scriptsize{}{}100 ns}{\scriptsize{}{} }  & \textcolor{black}{\scriptsize{}{}{}1.5 $\mu$s}{\scriptsize{}{}
}  & \textcolor{black}{\scriptsize{}{}120 $\mu$s}{\scriptsize{}{} }  & {\scriptsize{}{}{5 ns}}\tabularnewline
\hline 
{\scriptsize{}{}{{}Interaction time, $\tau_{i\mathrm{nt}}$} }  & {\scriptsize{}{}320 ps }  & {\scriptsize{}{}300 ps }  & {\scriptsize{}{}$\sim$1.43 ns }  & {\scriptsize{}{}440 ps}\tabularnewline
\hline 
{\scriptsize{}{}{{}Repetition rate, $R_{S}$} }  & {\scriptsize{}{}{1.2 GHz} }  & {\scriptsize{}{}{{}1.25 GHz} }  & {\scriptsize{}{}{{}518 MHz} }  & {\scriptsize{}{}{$\sim$667 MHz}}\tabularnewline
\hline 
\end{tabular}
\par\end{centering}

 {\scriptsize{}{}\protect\caption{\label{tab:warm-memories}{Properties of a selection of demonstrated
warm vapor memories. All values are derived from the corresponding
references given in the table. We denote the warm vapour (WV) memories of Refs. \cite{cavity_Raman_2016,Wal_mem,mag_shield,ORCA}
as WV1 through WV4, respectively. }}
}{\scriptsize \par}

\end{table}

\begin{figure}
\begin{centering}
\includegraphics[scale=0.3]{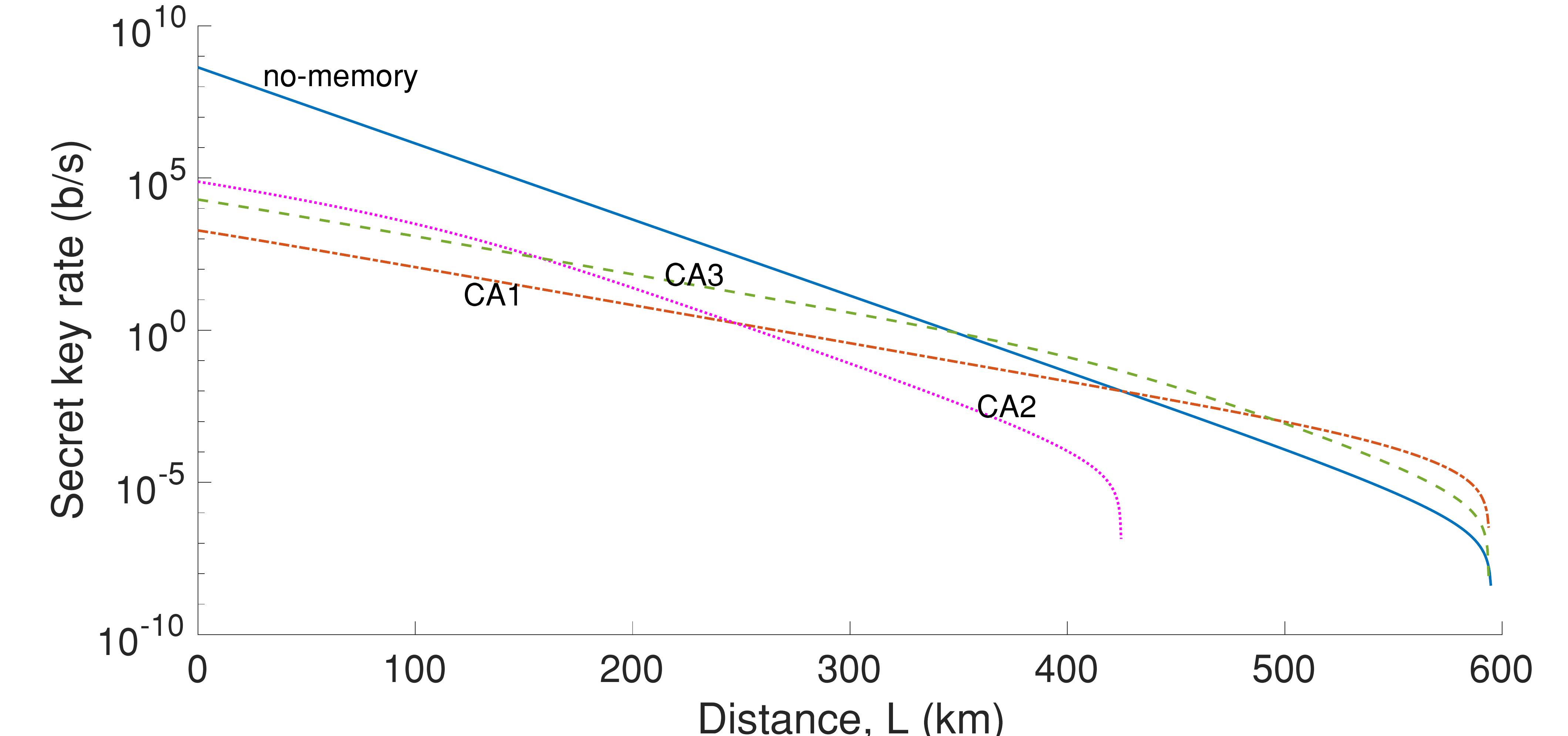} 
\par\end{centering}
\protect\caption{\label{fig:cold-memories}{Secret key rate for the setups of Fig
\ref{fig:setup_quasi} using the parameters of Table~\ref{tab:List-of-parameters}
and the cold atom memories featured in Table~\ref{tab:cold-memories}.
}}
\end{figure}

\emph{Cold atoms:} We consider the three experiments described in
Refs. \cite{cold1,cold2,cold3}. Reference \cite{cold2} utilizes
$^{85}$Rb in a magneto-optical trap while Refs. \cite{cold1,cold3}
feature atomic lattices of $^{87}$Rb. The coherence times of the
magneto-optical trap implementations are limited by, among many factors,
atomic diffusion in comparison to those of the atomic lattice \cite{cold1,cold2,cold3}. The
exceptional coherence time of Ref. \cite{cold1} is due to compensation
of light shifts, the insensitivity of the spin states to magnetic
field fluctuations, and use of dynamical decoupling. We note that
even though Ref. \cite{cold1} did not explicitly show storage of
non-classical light, the experiment of Ref. \cite{cold_DD} importantly shows
that no noise is introduced by dynamical decoupling. Our simulations, which are presented in Fig. \ref{fig:cold-memories}, show that
the atomic lattice experiments of Refs. \cite{cold1,cold3}
can allow rates that surpass the no-memory bound. Both memories have such long coherence times that, in both cases, the maximum security distance has been dictated by the dark count noise, and not the memory decoherence. However, these memories are only
useful if a low secret key rate is acceptable. Towards the possibility
of higher rates and shorter-distance operation, we consider small
improvements to memory properties (e.g. bandwidth) in Section C. Note
that the experiments of \cite{cold1} and \cite{cold3} employ off-resonant
Raman scattering to achieve memory-photon entanglement and have not
explicitly performed storage of an externally-provided photon as is
required for the quasi-EPR scheme. We assume that the QM parameters
derived from these experiments may be translated to a Raman memory
demonstration (as is achieved in \cite{cold2}). We also mention that
there is a theoretical proposal \cite{proposal} for Raman memory
using an optical lattice.

\begin{table}
\begin{centering}
{\scriptsize{}{}}%
\begin{tabular}{|c|c|c|c|}
\hline 
\textcolor{black}{\scriptsize{}{}{}}{\scriptsize{}{} }  & {\scriptsize{}{}CA1 \cite{cold1} }  & {\scriptsize{}{}CA2 \cite{cold2} }  & {\scriptsize{}{}CA3 \cite{cold3} }\tabularnewline
\hline 
\textcolor{black}{\scriptsize{}{}Efficiency, $\eta_{w}\eta_{r0}$}{\scriptsize{}{}
}  & {\scriptsize{}{}{{}0.14} }  & {\scriptsize{}{}{{}0.27} }  & {\scriptsize{}{}{{}0.76}}\tabularnewline
\hline 
{\scriptsize{}{}
}\textcolor{black}{\scriptsize{}{}{}Coherence time, $T_{r}$}{\scriptsize{}{}
}  & \textcolor{black}{\scriptsize{}{}16 s}{\scriptsize{}{} }  & \textcolor{black}{\scriptsize{}{}{}1.4 $\mu$s}{\scriptsize{}{}
}  & \textcolor{black}{\scriptsize{}{}220 ms}\tabularnewline
\hline 
{\scriptsize{}{}{{}Interaction time, $\tau_{i\mathrm{nt}}$} }  & {\scriptsize{}{}82 ns }  & {\scriptsize{}{}7 ns }  & {\scriptsize{}{}240 ns}\tabularnewline
\hline 
{\scriptsize{}{}{{}Repetition rate, $R_{S}$} }  & {\scriptsize{}{}{12 MHz} }  & {\scriptsize{}{}{133 MHz} }  & {\scriptsize{}{}{4.2 MHz}}\tabularnewline
\hline 
\end{tabular}
\par\end{centering}

 {\scriptsize{}{}\protect\caption{\label{tab:cold-memories}{Properties of a selection of demonstrated
cold atom memories and the corresponding interaction times and repetition
rates used for the numerical calculation of the secret key rate of
the quasi-EPR scheme. We denote the cold atom (CA) memories of Refs.
\cite{cold1,cold2,cold3} as CA1 through CA3, respectively. }}
}{\scriptsize \par}

\end{table}

\emph{Rare-earth-ion-doped crystals:} We consider the five AFC experiments
described in Refs. \cite{151Eu:Y2SiO5,153Eu:Y2SiO5cavity,Pr:Y2SiO5cavity,Pr:Y2SiO5,Pr:Y2SiO5non_classical}.
Europium-doped Y$_{2}$SiO$_{5}$ crystals are employed in the investigations
of Refs. \cite{151Eu:Y2SiO5,153Eu:Y2SiO5cavity} while the well-studied
Pr:Y$_{2}$SiO$_{5}$ is featured in Refs. \cite{Pr:Y2SiO5,Pr:Y2SiO5non_classical}.
On-demand storage at the single photon level is shown in Ref. \cite{151Eu:Y2SiO5},
of which dynamical decoupling techniques are also used to overcome
dephasing due to spin inhomogeneous broadening. Reference \cite{153Eu:Y2SiO5cavity}
utilizes a low-finesse cavity to show (up to 50\%) efficient and on-demand
storage of strong pulses. Efficient storage using a low-finesse cavity
is achieved in \cite{Pr:Y2SiO5cavity}, while on-demand storage of
qubits and heralded single photons are shown in Refs. \cite{Pr:Y2SiO5}
and \cite{Pr:Y2SiO5non_classical}, respectively. Again we simulate
the key rate of the quasi-EPR scheme and find that none of the REIC
implementations will surpass the no-memory performance. The best performance is offered by REIC2, which has a high efficiency and a decent coherence time. Taking into consideration the technical challenges of obtaining both high efficiency
and low noise in a REIC-based AFC system, in Section C we explore
the possibility of using several (spectral) modes to overcome the
no-memory bound. Note that coherence times of 6 hours \cite{6hT}
and one minute \cite{Pr_coherence} have been measured using magnetically-insensitive
ground-level transitions of $^{151}$Eu:Y$_{2}$SiO$_{5}$ and Pr:Y$_{2}$SiO$_{5}$,
respectively. However, it has yet to be shown that these coherence
times may be combined with the possibility of efficient and broadband
storage, hence these transitions may not be suitable for MA-MDI-QKD.
\begin{table}
\begin{centering}
{\scriptsize{}{}}%
\begin{tabular}{|c|c|c|c|c|c|}
\hline 
\textcolor{black}{\scriptsize{}{}{}}{\scriptsize{}{} }  & {\scriptsize{}{}REIC1 \cite{151Eu:Y2SiO5} }  & {\scriptsize{}{}REIC2 \cite{153Eu:Y2SiO5cavity} }  & {\scriptsize{}{}REIC3 \cite{Pr:Y2SiO5cavity} }  & {\scriptsize{}{}REIC4 \cite{Pr:Y2SiO5} }  & {\scriptsize{}{}REIC5 \cite{Pr:Y2SiO5non_classical}}\tabularnewline
\hline 
\textcolor{black}{\scriptsize{}{}Efficiency, $\eta_{w}\eta_{r0}$}{\scriptsize{}{}
}  & {\scriptsize{}{}{{}0.06} }  & {\scriptsize{}{}{{}0.53} }  & {\scriptsize{}{}{{}0.56}}  & {\scriptsize{}{}{{}0.04}}  & {\scriptsize{}{}{{}0.11}}\tabularnewline
\hline 
{\scriptsize{}{}
}\textcolor{black}{\scriptsize{}{}{}Coherence time, $T_{r}$}{\scriptsize{}{}
}  & \textcolor{black}{\scriptsize{}{}0.7 ms}{\scriptsize{}{} }  & \textcolor{black}{\scriptsize{}{}{}37 $\mu$s}{\scriptsize{}{}
}  & \textcolor{black}{\scriptsize{}{}3 $\mu$s}{\scriptsize{} } & \textcolor{black}{\scriptsize{}{}38 $\mu$s}{\scriptsize{}{} }  & \textcolor{black}{\scriptsize{}{}50 $\mu$s}\tabularnewline
\hline 
{\scriptsize{}{}{{}Repetition rate, $R_{S}$} }  & {\scriptsize{}{}{2 MHz} }  & {\scriptsize{}{}{5 MHz} }  & {\scriptsize{}{}{3 MHz} }  & {\scriptsize{}{}{3.5 MHz} }  & {\scriptsize{}{}{1 MHz}}\tabularnewline
\hline 
\end{tabular}
\par\end{centering}

 {\scriptsize{}{}\protect\caption{\label{tab:rare-earth-memories}{Properties of a selection of demonstrated
rare-earth-ion-doped memories and the corresponding interaction times
and repetition rates used for the numerical calculation of the secret
key rate of the quasi-EPR scheme. We denote the REIC memories of Refs.
\cite{151Eu:Y2SiO5,153Eu:Y2SiO5cavity,Pr:Y2SiO5cavity,Pr:Y2SiO5,Pr:Y2SiO5non_classical}
as REIC1 through REIC5, respectively.}}
}{\scriptsize \par}

\end{table}

\begin{figure}
\begin{centering}
\includegraphics[scale=0.3]{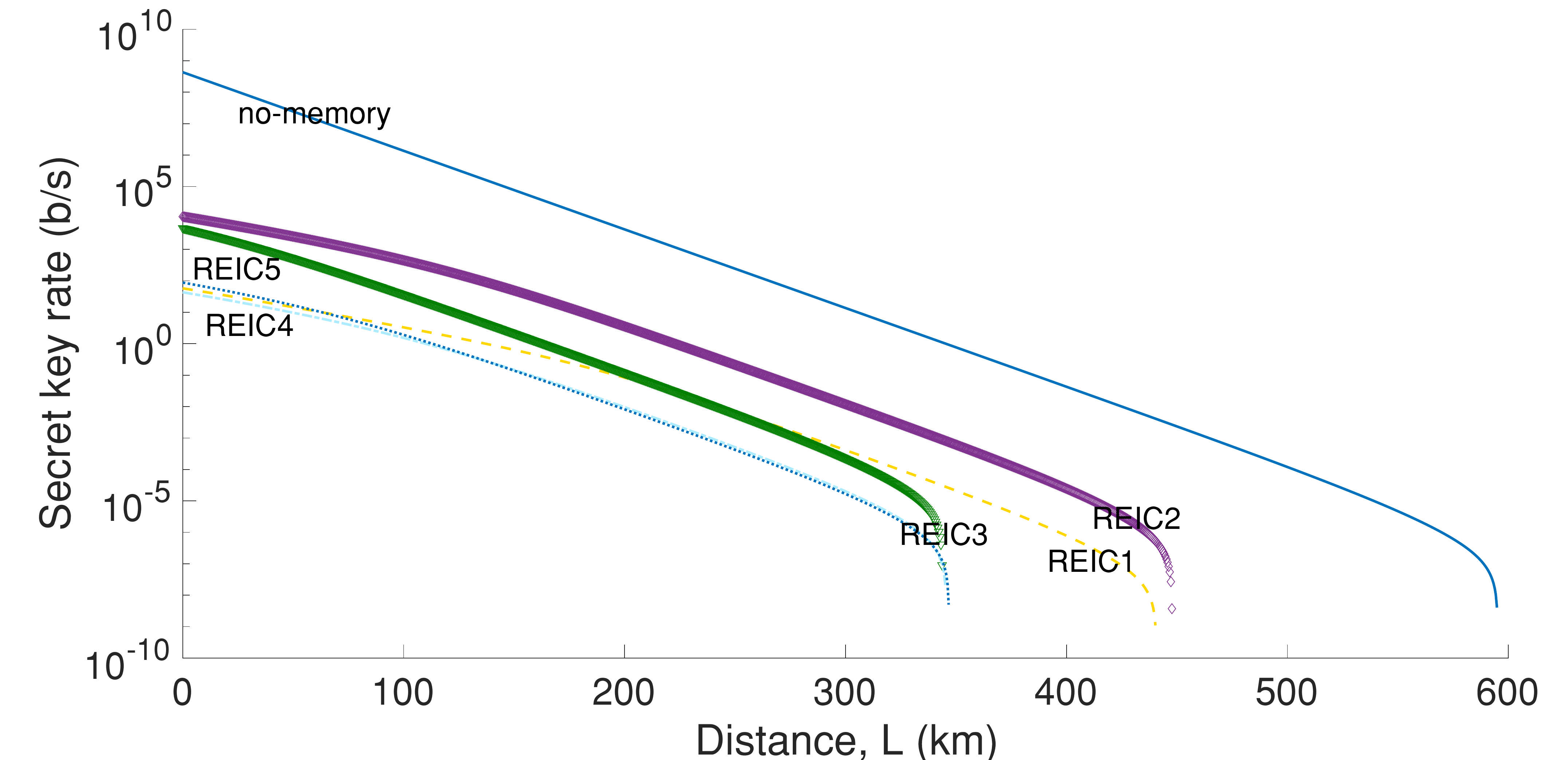} 
\par\end{centering}
\protect\caption{\label{fig:rare-earth-memories}{Secret key rate for the setups of
Fig \ref{fig:setup_quasi} using the parameters of Table~\ref{tab:List-of-parameters}
and the REIC memories featured in Table~\ref{tab:rare-earth-memories}.
}}
\end{figure}

\subsection{Near-future quantum memories}

In this section we evaluate the performance of the quasi-EPR scheme
using near-future QMs. Specifically, we suggest memory parameters
that could be obtained with realistic experimental improvements to
the memories of Refs. \cite{cavity_Raman_2016,Wal_mem,mag_shield,ORCA,cold1,cold2,cold3,151Eu:Y2SiO5,153Eu:Y2SiO5cavity,Pr:Y2SiO5cavity,Pr:Y2SiO5,Pr:Y2SiO5non_classical}.
We attempt to be conservative with our suggested parameters, in particular
with those of efficiency and coherence time, and acknowledge that
there are fundamental limitations of some parameters, e.g. the restriction
of bandwidth due to a certain energy level structure. Our enhanced
memory parameters may represent a short-term goal for developing QMs.

\begin{figure}
\begin{centering}
\includegraphics[scale=0.3]{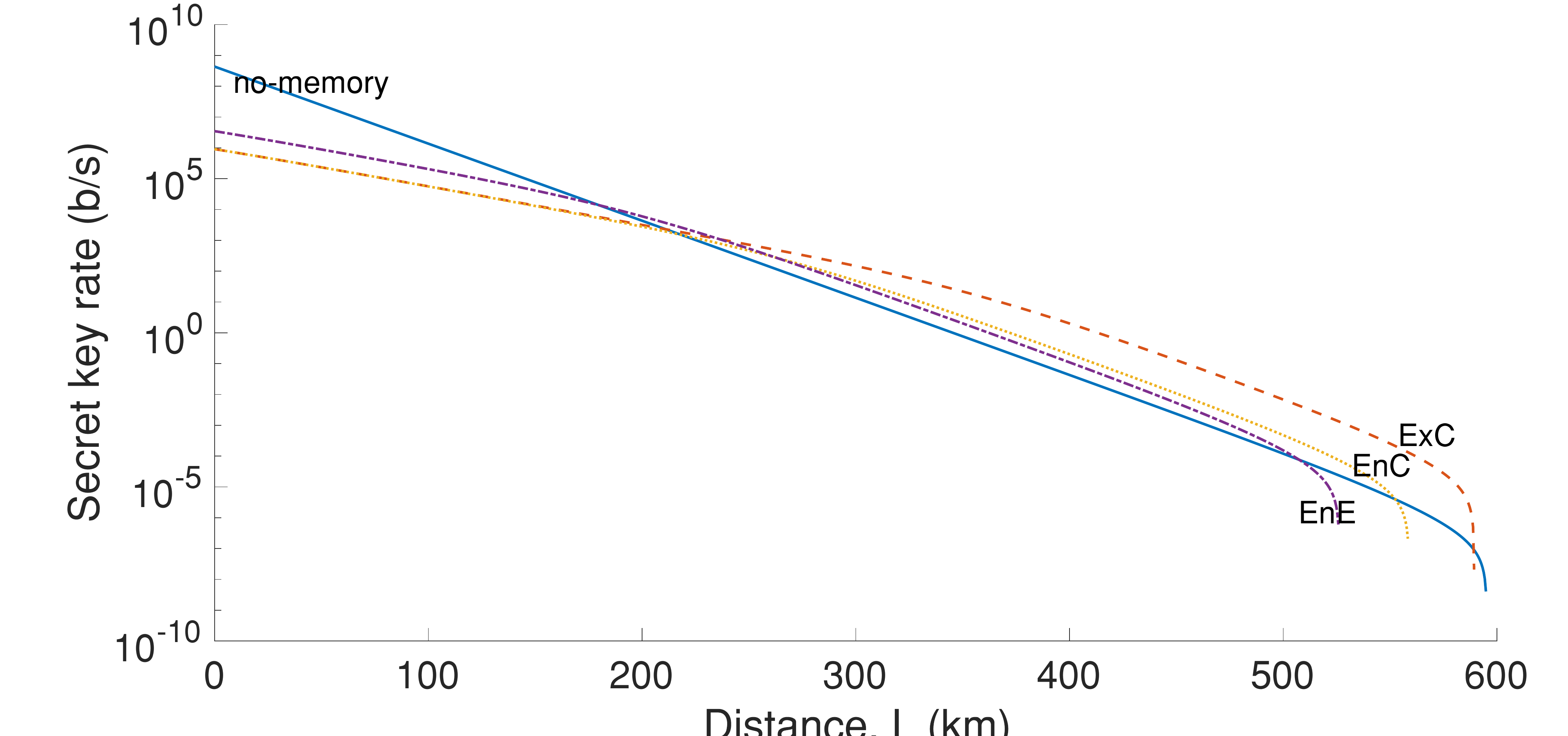} 
\par\end{centering}
\protect\caption{\label{fig:future-warm-memories}{ Secret key rate for the setups
of Fig \ref{fig:setup_quasi} using the parameters of Table~\ref{tab:List-of-parameters}
and the near-future warm vapor memories featured in Table~\ref{tab:warm-future-memories}.
}}
\end{figure}

\emph{Warm vapor:} Here we consider three potential QMs with
properties displayed in Table \ref{tab:warm-future-memories} and
corresponding quasi-EPR key rates shown in Fig. \ref{fig:future-warm-memories}.
The first we refer to as ``excellent coherence'' (ExC) in which improved
magnetic shielding will eliminate inhomogeneous spin dephasing such that a coherence time of Ref. \cite{mag_shield} is achieved.
Furthermore, we assume that a cavity is used to ensure low noise operation
\cite{cavity_Raman_2016} and an enhancement of efficiency to that
of Ref. \cite{Wal_mem}, either by the cavity or control field tailoring \cite{Raman_memory_theory}.
We find that this memory enables surpassing the bound at just over 200 km and
obtains maximal advantage at 400-500 km. This is a promising result
given that MDI-QKD has been demonstrated over 400~km \cite{MDI400km}\textemdash a
distance for which channel stabilization has been realized. The second
we refer to as ``enhanced coherence'' (EnC) in which we keep all parameters
the same as ExC except the coherence time, of which corresponds to
the minimum required to surpass the no-memory bound. Surprisingly,
we find that a (reasonable) coherence time of approximately 10 $\mu$s
will beat the bound at around 200 km, while the difference with memory ExC
lies in the rate-distance scaling at longer distances. The last QM
we refer to as ``enhanced efficiency'' (EnE) in which we keep the parameters
the same as in Ref. \cite{Wal_mem} except we find the minimum efficiency
to beat the bound, this being an efficiency of 60\% at a distance
of less than 200 km. Although it is likely that the EnE memory
is challenging to achieve without added noise, improvements in experimental
geometry in conjunction with control field optimization may reach
this requirement without any compromise to coherence time. The QM of Ref. \cite{ORCA} is not useful for MA-MDI-QKD due to
the limited coherence time (up to 100 ns) of the (excited) level used
for storage.

\begin{table}
\begin{centering}
{\scriptsize{}{}}%
\begin{tabular}{|c|c|c|c|}
\hline 
\textcolor{black}{\scriptsize{}{}{}}{\scriptsize{}{} }  & {\scriptsize{}{}ExC}  & {\scriptsize{}{}EnC }  & {\scriptsize{}{}EnE }\tabularnewline
\hline 
\textcolor{black}{\scriptsize{}{}Efficiency, $\eta_{w}\eta_{r0}$}{\scriptsize{}{}
}  & {\scriptsize{}{}{{}0.30} }  & {\scriptsize{}{}{{}0.30} }  & {\scriptsize{}{}{{}0.60}}\tabularnewline
\hline 
{\scriptsize{}{}
}\textcolor{black}{\scriptsize{}{}{}Coherence time, $T_{r}$}{\scriptsize{}{}
}  & \textcolor{black}{\scriptsize{}{}120 $\mu$s}{\scriptsize{}{} }  & \textcolor{black}{\scriptsize{}{}{}10 $\mu$s}{\scriptsize{}{}
}  & \textcolor{black}{\scriptsize{}{}1.5 $\mu$s}\tabularnewline
\hline 
{\scriptsize{}{}{{}Repitition rate, $R_{S}$} }  & {\scriptsize{}{}{1.25 GHz} }  & {\scriptsize{}{}{{}1.2 GHz} }  & {\scriptsize{}{}{{}1.2 GHz}}\tabularnewline
\hline 
\end{tabular}
\par\end{centering}

 {\scriptsize{}{}\protect\caption{\label{tab:warm-future-memories}{ Parameters for near-future warm
vapor memories, and the corresponding interaction times and repetition
rates, used for our numerical calculation of the secret key rate assuming
the setup of Fig.~\ref{fig:setup_quasi}. Memory abbreviations are
explained in the main text.}}
}{\scriptsize \par}

\end{table}

\begin{figure}
\begin{centering}
\includegraphics[scale=0.3]{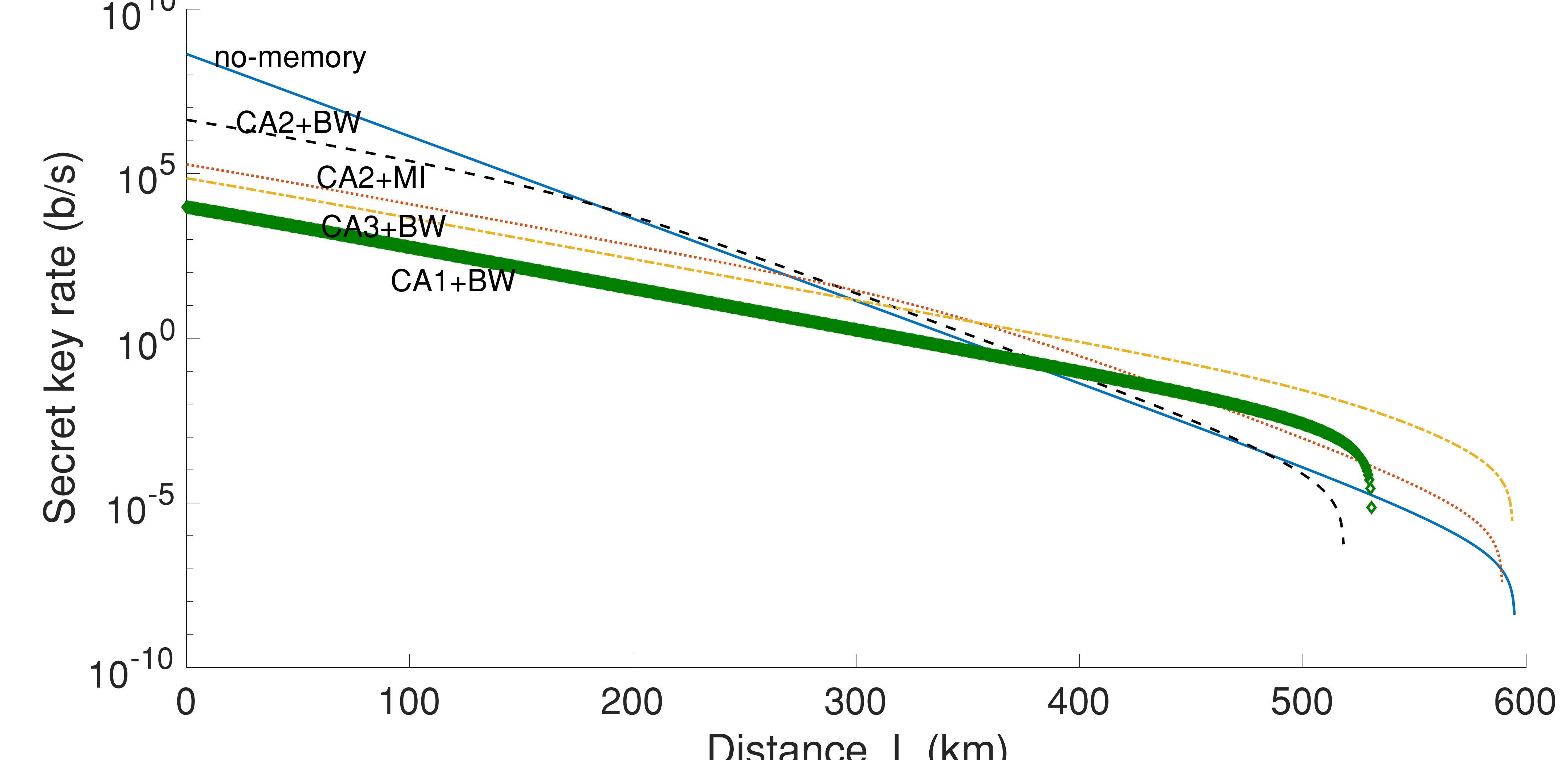} 
\par\end{centering}
\protect\caption{\label{fig:future-cold-memories}{ Secret key rate for the setups
of Fig \ref{fig:setup_quasi} using the parameters of Table~\ref{tab:List-of-parameters}
and the near-future cold atom memories featured in Table~\ref{tab:cold-future-memories}.
}}
\end{figure}

\emph{Cold atoms:} Here we consider the QMs outlined in Table
\ref{tab:cold-future-memories}, with the corresponding key rates
shown in Fig.~\ref{fig:future-cold-memories}. We consider the memory
of \cite{cold2} with a bandwidth expanded to 1 GHz (CA2+BW), of which
results in $R_{S}\sim667$ MHz. Note that the bandwidth must be less
than half of the 3 GHz ground-state splitting of $^{85}$Rb to ensure
minimum impact of noise. Unfortunately, we find that, due its low
coherence time, this QM will only (just) beat the no-memory bound if
it is $\sim$90\% efficient. Next we assume that a magnetically-insensitive
ground-state transition is employed for the investigation of \cite{cold2}
(CA2+MI), finding that about 50\% efficiency is needed to beat the bound, which can
be realized by control pulse shaping or backwards retrieval \cite{Raman_memory_theory}.
We also consider the QM of Ref. \cite{cold3} except we allow
the bandwidth to be expanded to 100 MHz, $R_{S}=95$ MHz, (CA3+BW)
which is well below the limitations given by the ground-state structure,
but may pose a challenge if a cavity setup is employed. Encouragingly,
we find that this QM easily overcomes the bound if it is 30\%
efficient. Finally, if the highly-coherent memory of
Ref. \cite{cold1} is employed and its bandwidth is expanded from 12.2 to
100 MHz (CA1+BW), only 10\% efficiency is required to be useful for
MDIQKD for distances greater than 600 km, albeit at a low key rate. Note that, in majority of cases, the cross-over distance is around 300~km.

\begin{table}
\begin{centering}
{\scriptsize{}{}}%
\begin{tabular}{|c|c|c|c|c|}
\hline 
\textcolor{black}{\scriptsize{}{}{}}{\scriptsize{}{} }  & {\scriptsize{}{}CA2+BW}  & {\scriptsize{}{}CA2+MI}  & {\scriptsize{}{}CA3+BW } & {\scriptsize{}{}CA1+BW }\tabularnewline
\hline 
\textcolor{black}{\scriptsize{}{}Efficiency, $\eta_{w}\eta_{r0}$}{\scriptsize{}{}
}  & {\scriptsize{}{}{{}0.90} }  & {\scriptsize{}{}{{}0.50} }  & {\scriptsize{}{}{{}0.30}} & {\scriptsize{}{}{{}0.10} }\tabularnewline
\hline 
{\scriptsize{}{}
}\textcolor{black}{\scriptsize{}{}{}Coherence time, $T_{r}$}{\scriptsize{}{}
}  & \textcolor{black}{\scriptsize{}{}1.4 $\mu$s}{\scriptsize{}{} }  & \textcolor{black}{\scriptsize{}{}{}1 ms}{\scriptsize{}{} }  & \textcolor{black}{\scriptsize{}{}220 ms} & \textcolor{black}{\scriptsize{}{}16 s}\tabularnewline
\hline 
{\scriptsize{}{}{{}Repetition rate, $R_{S}$} }  & {\scriptsize{}{}{$\sim$667 MHz} }  & {\scriptsize{}{}{95 MHz} }  & {\scriptsize{}{}{95 MHz}}& {\scriptsize{}{}{95 MHz} } \tabularnewline
\hline 
\end{tabular}
\par\end{centering}

 {\scriptsize{}{}\protect\caption{\label{tab:cold-future-memories}{ Parameters of near-future cold
atom memories, and the corresponding interaction times and repetition
rates, used for our numerical calculation of the secret key rate assuming
the setup of Fig.~\ref{fig:setup_quasi}. Memory abbreviations are
explained in the main text.}}
}{\scriptsize \par}

\end{table}

\begin{figure}
\begin{centering}
\includegraphics[scale=0.3]{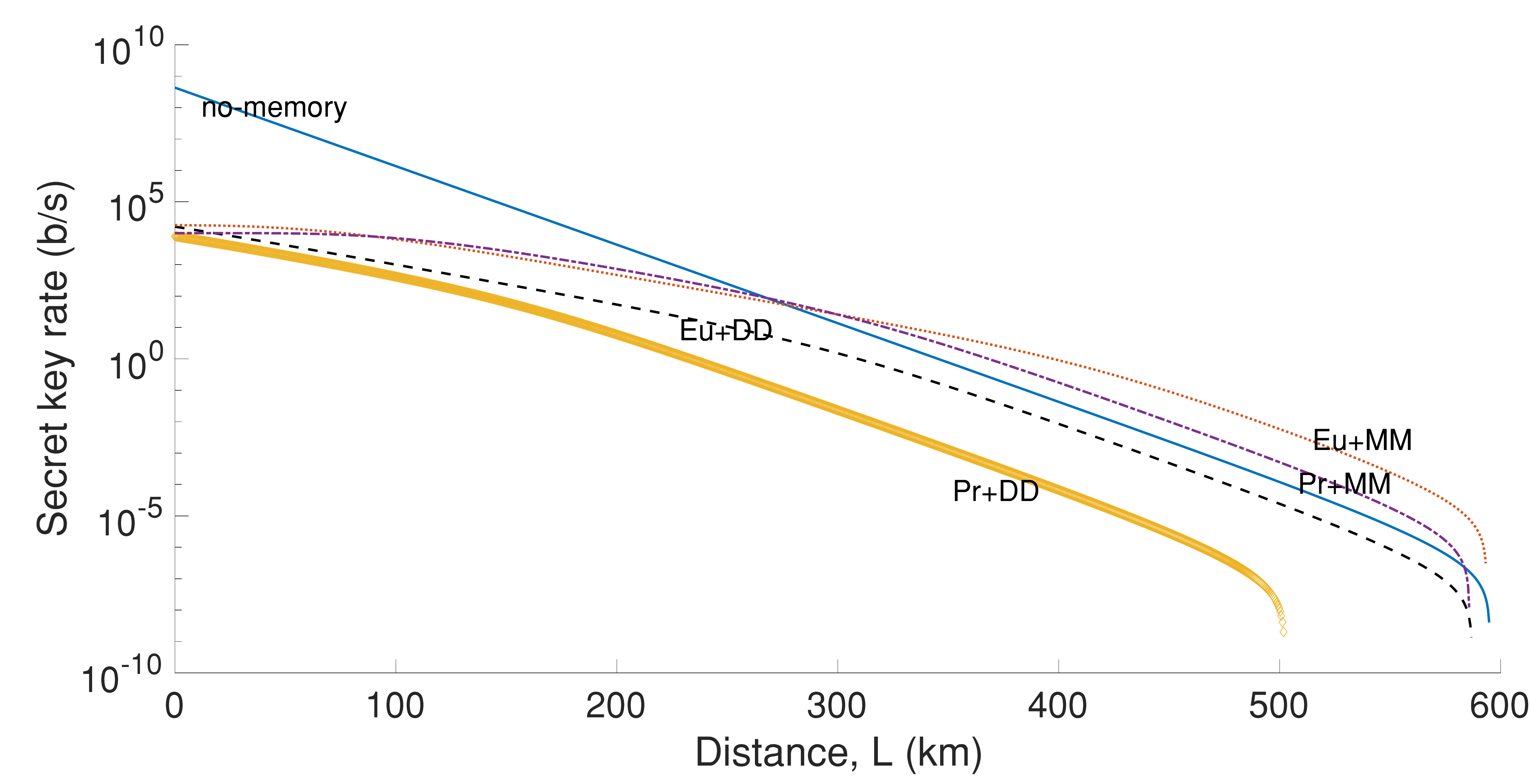} 
\par\end{centering}
\protect\caption{\label{fig:future-rare-earth-memories}{ Secret key rate for the
setups of Fig \ref{fig:setup_quasi} using the parameters of Table~\ref{tab:List-of-parameters}
and the REIC memories featured in Table~\ref{tab:rare-earth-future-memories}.
}}
\end{figure}

\emph{Rare-earth-ion-doped crystals:} The corresponding QM properties
and key rates are shown in Table \ref{tab:rare-earth-future-memories}
and Fig. \ref{fig:future-rare-earth-memories}, respectively. We employ
the $^{151}$Eu:Y$_{2}$SiO$_{5}$ memory of Ref. \cite{151Eu:Y2SiO5},
except that we assume perfect dynamical decoupling is in use to achieve
a coherence time that is entirely limited by the ground-level homogeneous
broadening (Eu+DD), and we employ the Pr:Y$_{2}$SiO$_{5}$ crystal
of \cite{Pr:Y2SiO5} and \cite{Pr:Y2SiO5non_classical} (Pr+DD) in
a similar way. Even with perfect efficiency, we find that neither of the QMs overcome the bound, mainly due to their limited bandwidth
in comparison to the Raman QMs. To gain an advantage, we first
assume the cavity enhanced setups of \cite{153Eu:Y2SiO5cavity} and
\cite{Pr:Y2SiO5cavity} in conjunction with memories Eu+DD and Pr+DD,
respectively. We then consider the possibility of multi-mode storage
and find the minimum number of modes that are needed to beat the bound,
which we refer to as memories Eu+MM and Pr+MM for Eu- and Pr-doped
Y$_{2}$SiO$_{5}$, respectively. Since our implementation is already
intrinsically temporally multi-mode, a convenient degree of freedom
to use for multiplexing is that of frequency. This is especially true
of REICs where their sub-level structure limits the AFC bandwidth,
but their inhomogeneously-broadened lines offer simultaneous storage of many, in some cases up to one-thousand \cite{AFC_multi}, spectral modes \cite{macfarlane1987,thiel_review}. Praseodymium-doped Y$_{2}$SiO$_{5}$
offers the possibility to store up to $\sim$100 spectral modes given
its hyperfine structure and its $\sim$5 GHz inhomogeneous linewidth
\cite{Pr:Y2SiO5}, while $^{151}$Eu:Y$_{2}$SiO$_{5}$ only offers
the possibility of storing a single spectral mode \cite{151Eu:Y2SiO5}.
Nonetheless, one could employ spatial multiplexing, or explore the
possibility to increase the inhomogneous linewidth by co-doping methods
\cite{Sun2005}. In order to use the multi-mode feature of the memory we may need to employ an array of SPSs, each generating single photons at different wavelength or spatial modes. One should account for that if the normalized rate per channel use is of interest.


\begin{table}
\begin{centering}
{\scriptsize{}{}}%
\begin{tabular}{|c|c|c|c|c|}
\hline 
\textcolor{black}{\scriptsize{}{}{}}{\scriptsize{}{} }  & {\scriptsize{}{}Eu+DD}  & {\scriptsize{}{}Eu+MM }  & {\scriptsize{}{}Pr+DD }  & {\scriptsize{}{}Pr+MM}\tabularnewline
\hline 
\textcolor{black}{\scriptsize{}{}Efficiency, $\eta_{w}\eta_{r0}$}{\scriptsize{}{}
}  & {\scriptsize{}{}{{}1} }  & {\scriptsize{}{}{{}0.53} }  & {\scriptsize{}{}{{}1}}  & {\scriptsize{}{}{{}0.56}}\tabularnewline
\hline 
{\scriptsize{}{}
}\textcolor{black}{\scriptsize{}{}{}Coherence time, $T_{r}$}{\scriptsize{}{}
}  & \textcolor{black}{\scriptsize{}{}15 ms}{\scriptsize{}{} }  & \textcolor{black}{\scriptsize{}{}{}15 ms}{\scriptsize{}{} }  & \textcolor{black}{\scriptsize{}{}500 $\mu$s}{\scriptsize{} } & \textcolor{black}{\scriptsize{}{}500 $\mu$s}\tabularnewline
\hline 
{\scriptsize{}{}{{}Repitition rate, $R_{S}$} }  & {\scriptsize{}{}{2 MHz} }  & {\scriptsize{}{}{2 MHz} }  & {\scriptsize{}{}{1 MHz} }  & {\scriptsize{}{}{1 MHz}}\tabularnewline
\hline 
{\scriptsize{}{}{{}Number of spectral modes, $N$} }  & {\scriptsize{}{}{{}1} }  & {\scriptsize{}{}{{}30} }  & {\scriptsize{}{}{{}1} }  & {\scriptsize{}{}{{}90}}\tabularnewline
\hline 
\end{tabular}
\par\end{centering}

 {\scriptsize{}{}\protect\caption{\label{tab:rare-earth-future-memories}{ Parameters of near-future
rare-earth-ion-doped memories, and the corresponding interaction times
and repetition rates, used for our numerical calculation of the secret
key rate assuming the setup of Fig.~\ref{fig:setup_quasi}. Memory
abbreviations are explained in the main text.}}
}{\scriptsize \par}

\end{table}

\subsection{Near-future memories with additional system imperfections}

\label{Sec:additional} The implication that atomic ensembles could
outperform the no-QM QKD is based on several assumptions.
The key assumption is that the two SPSs in the quasi-EPR setup can
generate identical single photons that are (bandwidth-) matched to the QMs. We have also
thus far ignored the additional background noise coming from the frequency
converters. Any deviation from these assumptions may change the rate
scaling and add to the QBER of the system. Below, we use our rough
calculations of Sec.~\ref{sec:QEPR} to investigate how resilient our setup
is to the following imperfections. 
\begin{itemize}
\item \textbf{Multi-photon terms:} We now test the resilience of our setup against possible multiple-photon components in the SPS. In fact,
one can say that so long as $p_{2}/p_{1}\ll\exp(-L/(2L_{{\rm att}}))$,
our system is immune against the two-photon terms generated by the source.
At $L=200$~km, that would require $p_{2}/p_{1}\ll0.003$, which
is almost achievable with today's quantum dot technology for generating
entangled and/or single photons \cite{QDot_entg_low_g2:NatPhot2014},
and possibly even those that rely on parametric down-conversion. In
the latter case, a bank of downconverters is needed to boost the trigger
rate of the system \cite{SPDC_multiplex}. The additional QBER due to two-photon terms is
also on the order of $p_{2}$, which is negligible. 
\item \textbf{Photons distinguishibility:} If the two single photons generated
by the two SPSs in Fig.~\ref{fig:setup_quasi}(b) do not fully couple
to each other at 50:50 beam splitters, then some TLIC-related issues occur at the side BSMs. Yet, similar to the two-photon terms, our system can tolerate
the same order of magnitude (0.1\textendash 1\%) mismatch between
the corresponding modes of the two single photons, which is again
achievable by the current technology \cite{SomaschiN.2016}. The additional
QBER is also expected to be on the same order. The overlap between
the user's photon and the SPSs in the middle node is important, but
not as vital as the overlap between that of the two SPSs. The former
issue could increase the QBER to some extent but given that long-distance
MDI-QKD has been demonstrated, this issue can be dealt with using
existing technologies. 
\item \textbf{Bandwidth Mismatch:} If the bandwidth of the SPS photons and
the QM do not match, one may end up with a large loss factor in the
writing efficiency. For instance, the bandwidth of cold atomic ensembles
is on the order of 10-100 MHz, which does not match that of many
quantum-dot sources. If a quantum-dot source is used
with CA1\textendash CA3 memories, a drop of one to two orders of magnitude
may be expected in their corresponding key rates in Figs.~\ref{fig:cold-memories}
and \ref{fig:future-cold-memories}. The situation is more promising
for warm vapor QMs, as their bandwidths are compatible with that of quantum-dot
sources.

\item \textbf{Background photons:} Finally, we have thus far ignored the effect of additional background noise generated by the frequency converters in our numerical analysis. In principle, at $L=200$~km, based
on the condition $d_{c}\ll\exp(-L/(2L_{{\rm att}}))$, one expects to
tolerate a dark count on the order of $10^{-4}$ per pulse, which
is an order of magnitude higher than the typical background noise
from frequency converters \cite{Upconversion_1550to800}. 
\end{itemize}
\begin{figure}
\begin{centering}
\includegraphics[width=0.7\textwidth]{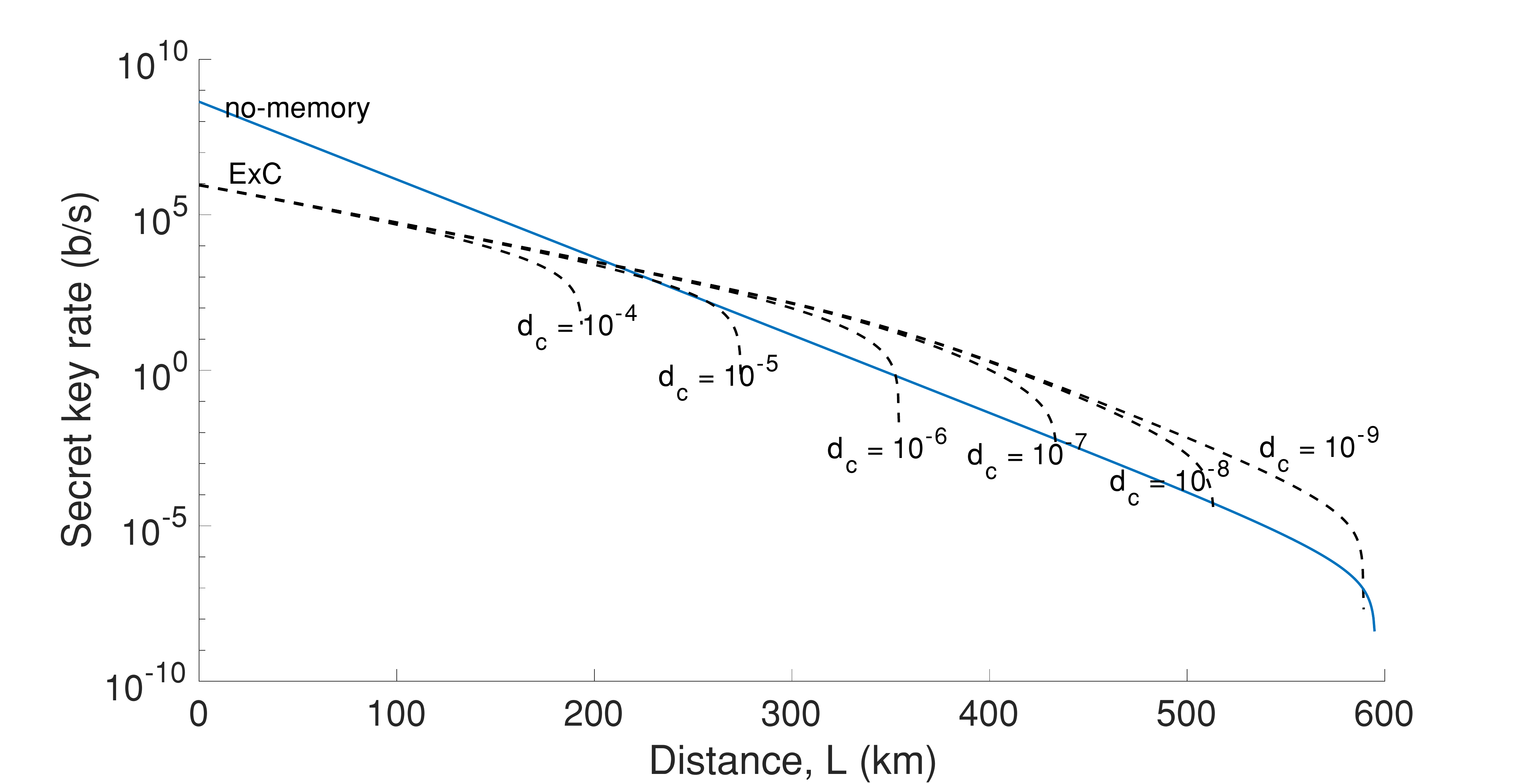}
\par\end{centering}
\protect\caption{\label{fig:rate_dc}{\footnotesize{}{}{}Comparison of the secret
key rates of the setup of Fig. \ref{fig:setup_quasi} with near-future warm vapor
atomic ensembles at different values of dark count for the side-BSM
detectors. The dark count term here accounts for not only the detector dark count, but also the background noise due to frequency converters and possibly the QMs. The other values are as in memory ExC in Tables~\ref{tab:warm-future-memories}
and \ref{tab:List-of-parameters}. }}
\end{figure}

In order to test the above expectations, in Fig.~\ref{fig:rate_dc},
we have plotted the effect of dark counts from the side-BSM modules on
the key rate of the MA-QKD system that uses memory ExC from near-future
warm vapor atomic ensembles. Since warm vapor QMs are employed, no loss due to bandwidth mismatch is considered. The results show that, at $d_{c}=10^{-6}$, the rate is nearly one order of magnitude above the MDI-QKD curve at $L=300$~km, which leaves room for losses due to other experimental imperfections. Note that such a study of dark noise also guides the development of future Raman QMs based on warm vapor of which, without special considerations, are plagued by four-wave-mixing-induced noise \cite{cavity_Raman_2016}.

\section{Conclusions}

In this paper we explored the possibility of using ensemble-based QMs in MA-MDI-QKD setups. Such QMs promise high efficiencies due to their strong light-matter coupling, large time-bandwidth products, and the ability to store multiple modes. By using single-photon sources, which are at an advanced stage of development, we proposed setups that could remove or alleviate the (single-mode) multiple-excitation problem. We identified the key problems in previously-proposed setups or the ones that resembled NLAs, and proposed a quasi-EPR setup that could outperform single no-memory QKD links. 
We showed that our solution is resilient against main imperfections
in the source, the QM module, and other required devices such
as frequency converters and single-photon detectors. Based on our
calculations, warm vapor atomic ensembles have the best chance to
improve the rate-versus-distance behavior at channel distances above
200~km provided their efficiencies and coherence times can be improved.
Cold atomic ensembles also offer a good performance provided that
the bandwidth mismatch between the QMs and the driving SPSs can
be reduced. Certain AFC memories, such as {{}$^{153}\mathrm{Eu}^{3+}\mathrm{:Y}_{2}\mathrm{SiO}_{5}$},
were also able to get close to the no-QM systems, but they need improvement
in their coupling efficiency, coherence time and multi-mode capacity
in order to offer a stable improvement. Our analysis ensures that
a proof-of-principle experiment for our proposed setup would be within
reach of the current technology and sets the stage for larger
quantum repeater links to be implemented in the future.

\section*{Acknowledgments}
This work was partly funded by the UK's EPSRC Grant EP/M013472/1 and EPSRC
Grant EP/M506951/1, and the EU's H2020 programme under the Marie Sk\l{}odowska-Curie project QCALL (GA 675662).

\bibliographystyle{apsrev4-1}
\bibliography{bib1}

\end{document}